\documentclass[reprint, nofootinbib, prb, superscriptaddress]{revtex4-1}
\usepackage{amsmath} \usepackage{graphicx} \usepackage{dcolumn}
\usepackage{bm} 
\usepackage{amssymb}
\usepackage{pstricks}

\begin{document}

\newlength{\figurewidth}
\setlength{\figurewidth}{\columnwidth}

\newcommand{\prtl}{\partial}
\newcommand{\la}{\left\langle}
\newcommand{\ra}{\right\rangle}
\newcommand{\dla}{\la \! \! \! \la}
\newcommand{\dra}{\ra \! \! \! \ra}
\newcommand{\we}{\widetilde}
\newcommand{\smfp}{{\mbox{\scriptsize mfp}}}
\newcommand{\smp}{{\mbox{\scriptsize mp}}}
\newcommand{\sph}{{\mbox{\scriptsize ph}}}
\newcommand{\sinhom}{{\mbox{\scriptsize inhom}}}
\newcommand{\sneigh}{{\mbox{\scriptsize neigh}}}
\newcommand{\srlxn}{{\mbox{\scriptsize rlxn}}}
\newcommand{\svibr}{{\mbox{\scriptsize vibr}}}
\newcommand{\smicro}{{\mbox{\scriptsize micro}}}
\newcommand{\scoll}{{\mbox{\scriptsize coll}}}
\newcommand{\sattr}{{\mbox{\scriptsize attr}}}
\newcommand{\sth}{{\mbox{\scriptsize th}}}
\newcommand{\sauto}{{\mbox{\scriptsize auto}}}
\newcommand{\seq}{{\mbox{\scriptsize eq}}}
\newcommand{\teq}{{\mbox{\tiny eq}}}
\newcommand{\sinn}{{\mbox{\scriptsize in}}}
\newcommand{\suni}{{\mbox{\scriptsize uni}}}
\newcommand{\tin}{{\mbox{\tiny in}}}
\newcommand{\scr}{{\mbox{\scriptsize cr}}}
\newcommand{\tstring}{{\mbox{\tiny string}}}
\newcommand{\sperc}{{\mbox{\scriptsize perc}}}
\newcommand{\tperc}{{\mbox{\tiny perc}}}
\newcommand{\sstring}{{\mbox{\scriptsize string}}}
\newcommand{\stheor}{{\mbox{\scriptsize theor}}}
\newcommand{\sGS}{{\mbox{\scriptsize GS}}}
\newcommand{\sBP}{{\mbox{\scriptsize BP}}}
\newcommand{\sNMT}{{\mbox{\scriptsize NMT}}}
\newcommand{\sbulk}{{\mbox{\scriptsize bulk}}}
\newcommand{\tbulk}{{\mbox{\tiny bulk}}}
\newcommand{\sXtal}{{\mbox{\scriptsize Xtal}}}
\newcommand{\sliq}{{\text{\tiny liq}}}

\newcommand{\smin}{\text{min}}
\newcommand{\smax}{\text{max}}

\newcommand{\saX}{\text{\tiny aX}}
\newcommand{\slaX}{\text{l,{\tiny aX}}}

\newcommand{\svap}{{\mbox{\scriptsize vap}}}
\newcommand{\sjam}{J}
\newcommand{\Tm}{T_m}
\newcommand{\sTS}{{\mbox{\scriptsize TS}}}
\newcommand{\sDW}{{\mbox{\tiny DW}}}
\newcommand{\cN}{{\cal N}}
\newcommand{\cB}{{\cal B}}
\newcommand{\br}{\bm r}
\newcommand{\be}{\bm e}
\newcommand{\cH}{{\cal H}}
\newcommand{\cHlt}{\cH_{\mbox{\scriptsize lat}}}
\newcommand{\sthermo}{{\mbox{\scriptsize thermo}}}

\newcommand{\bu}{\bm u}
\newcommand{\bk}{\bm k}
\newcommand{\bX}{\bm X}
\newcommand{\bY}{\bm Y}
\newcommand{\bA}{\bm A}
\newcommand{\bb}{\bm b}

\newcommand{\lintf}{l_\text{intf}}

\newcommand{\DV}{\delta V_{12}}
\newcommand{\sout}{{\mbox{\scriptsize out}}}
\newcommand{\dv}{\Delta v_{1 \infty}}
\newcommand{\dvin}{\Delta v_{2 \infty}}

\newcommand{\wtp}{\tilde{p}}
\newcommand{\wtK}{\widetilde{K}}
\newcommand{\wtgm}{\tilde{\gamma}}
\newcommand{\wtg}{\widetilde{g}}

\newcommand{\tU}{{\tiny U}}

\def\Xint#1{\mathchoice
   {\XXint\displaystyle\textstyle{#1}}%
   {\XXint\textstyle\scriptstyle{#1}}%
   {\XXint\scriptstyle\scriptscriptstyle{#1}}%
   {\XXint\scriptscriptstyle\scriptscriptstyle{#1}}%
   \!\int}
\def\XXint#1#2#3{{\setbox0=\hbox{$#1{#2#3}{\int}$}
     \vcenter{\hbox{$#2#3$}}\kern-.5\wd0}}
\def\ddashint{\Xint=}
\def\dashint{\Xint-}
\title{Temperature-driven narrowing of the insulating gap as a
  precursor of the insulator-to-metal transition: Implications for the
  electronic structure of solids}

\author{Vassiliy Lubchenko} \email{vas@uh.edu}
\affiliation{Department of Chemistry, University of Houston,
  Houston, TX 77204-5003} \affiliation{Department of Physics,
  University of Houston, Houston, TX 77204-5005}

\author{Arkady Kurnosov}\affiliation{Department of
  Chemistry, University of Houston, Houston, TX 77204-5003}

\date{\today}

\begin{abstract}

  We present a microscopic picture rationalizing the surprisingly
  steep decrease of the band gap with temperature in insulators,
  crystalline or otherwise. The gap narrowing largely results from
  fluctuations of long-wavelength optical phonons---when the latter
  are present---or their disordered analogs, if the material is
  amorphous.  We elaborate on this notion to show that possibly with
  the exception of weakly bound solids made of closed-shell atoms, the
  existence of an insulating gap or pseudo-gap in a periodic solid
  implies that optical phonons must be present, too. This means that
  in an insulating solid, the primitive cell must have at least two
  atoms and/or that a charge density wave is present.  As a corollary,
  a (periodic) elemental solid whose primitive unit contains only one
  atom will ordinarily be a metal, possibly unless the element belongs
  to group 18 of the periodic table, consistent with observation.
  Some implications of the present results for quantum solids are
  briefly discussed, such as that the ground state of the Wigner
  crystal must be anti-ferromagnetic.  A simple field theory of the
  metal-insulator transition is constructed that ties long-wavelength
  optical vibrations with fluctuations of an order parameter for the
  metal-insulator transition; symmetry-breaking aspects of the latter
  transition are thereby highlighted.

\end{abstract}

\maketitle

\section{Motivation}
\label{motivation}

\begin{figure}[t!]
\center
\includegraphics[width = .85 \figurewidth]{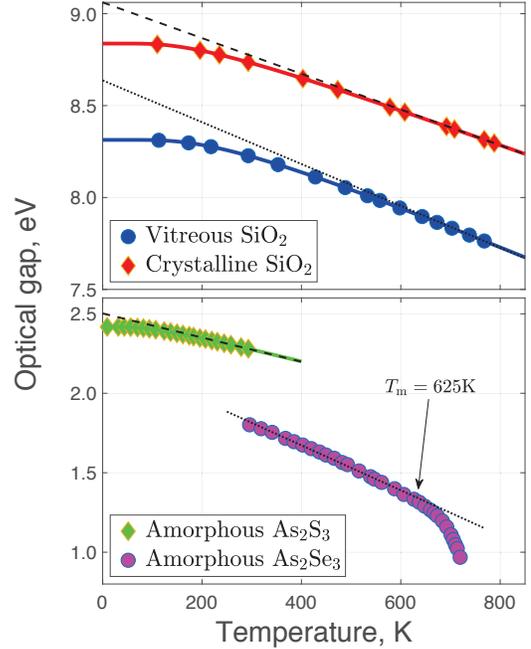}
\caption{\label{EgTFig} Optical gaps in vitreous and crystalline
  SiO$_{2}$ (top), and amorphous As$_{2}$S$_{3}$ and As$_{2}$Se$_{3}$
  (bottom) as functions of temperature. The quartz data were inferred
  from Ref.~\onlinecite{doi:10.1002/pssb.2221160133}, As$_{2}$S$_{3}$
  from Ref.~\onlinecite{0022-3719-7-8-022} and amorphous
  As$_{2}$Se$_{3}$ from Ref.~\onlinecite{ARAI1975295}. The choice of
  the respective absorption coefficients, listed in
  Table~\ref{VTable}, is explained in Section~\ref{empirics}. The
  solid lines represent fits obtained using the present description,
  see Section~\ref{optical}. The corresponding parameters are also
  listed in Table~\ref{VTable}. The straight lines correspond with the
  high-$T$ behavior from Eq.~(\ref{EgTlinear}).}
\end{figure}

The temperature-dependence of the insulating gap, which is exemplified
by the data for a small number of crystalline and amorphous compounds
in Fig.~\ref{EgTFig}, is significant for quantitative descriptions of
the electrical conductivity and many other electronic phenomena.  In a
substantial temperature interval, the dependence can be described by a
simple linear law:
\begin{equation} \label{EgTlinear} E_g \approx E_{g, 0} - \gamma T,
\end{equation}
where $E_{g, 0}$ and $\gamma$ are numerical constants.  The absorption
edge is not sharp even in crystalline compounds because optical
transitions are much faster than atomic motions; the instantaneous
positions of the ions are disordered already owing to thermal motions,
let along when the solid is amorphous. The so smeared optical edge is
often called the Urbach~\cite{PhysRev.92.1324, MARTIENSSEN1957257}
tail (or edge), or Lifshitz tail.~\cite{doi:10.1080/00018736400101061}

\begin{figure}[t]
\center
\includegraphics[width =  \figurewidth]{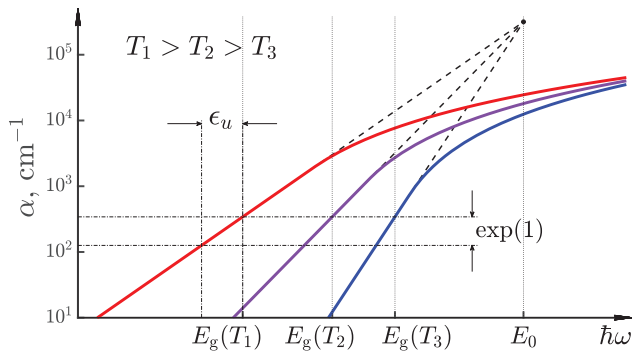}
\caption{\label{alphaCross} Typical behavior of the optical absorption
  spectrum as a function of temperature. $\alpha$ is the standard
  absorption coefficient. The width of the exponential tail of
  localized states is seen to change synchronously with the mobility
  edge.}
\end{figure}

Because of the edge smearing, one customarily defines the optical gap
at a given value of the absorption coefficient $\alpha$, the smaller
the $\alpha$ the greater the coefficient $\gamma$ from
Eq.~(\ref{EgTlinear}). This is illustrated in Fig.~\ref{alphaCross};
the original data can be found in
Refs~\onlinecite{doi:10.1002/pssb.2221160133, PhysRevLett.47.1480,
  PhysRev.92.1324, MARTIENSSEN1957257}. We show the values of the
coefficient $\gamma$ that are pertinent to activated charge transport;
these correspond to optical transitions between ``borderline'' states
that separate delocalized from localized states in the respective
bands. This will be discussed in more detail in the bulk of the paper.

The pertinent values of $\gamma$ are surprisingly large, as in
Fig.~\ref{EgTFig}. They imply that temperature changes on the order of
a few hundredths of eV cause changes in the electronic structure that
are an order of magnitude larger.  In addition to raising mechanistic
questions, this strong effect has direct consequences for estimating
the (temperature dependent) electrical conductivity $\sigma(T)$ in the
thermally activated regime.  Because the effective activation energy
for charge transport could be as large as $E_g/2$, the $(-\gamma T)$
term in Eq.~(\ref{EgTlinear}) could lead to an effective increase of
the prefactor in the activation rate by as much as $e^{\gamma/2}$, the
latter quantities listed in Table~\ref{VTable}. The actual increase is
usually smaller because only one of the carriers dominates; still it
is usually on the order of $10^2$, as will be discussed in a separate
submission.  Ignoring such a large factor could throw off one's
estimates of the density of states of charge carriers and/or their
mobility by an amount that would be difficult to write off on one's
being agnostic as to the detailed shape of the electronic density of
states or the precise strength of electron-lattice coupling.

Many substances exhibit a remarkable behavior, exemplified in
Fig.~\ref{alphaCross}, where the temperature-driven narrowing of the
band and the broadening of the exponential tail of localized states
occur in a synchronous fashion. More specifically, the onset energy of
the tail of the localized states is exactly proportional to the width
of the tail.~\cite{Cody1984} This remarkable behavior may or may not
be fully universal; at least in some cases deviations can be accounted
for by the difficulty in isolating the amount of midgap absorption due
to impurities.~\cite{doi:10.1002/pssb.2220640116}


Existing views on the microscopic origin of the temperature-driven
narrowing of the band gap seem to hark back to the work of
Fan~\cite{PhysRev.78.808.2, PhysRev.82.900} dating to the early
1950s. For polar substances, Fan co-opts Fr\"ohlich's
formalism~\cite{Frohlich230, PhysRev.76.1394} for the electron-lattice
coupling in ionic substances to conclude that in the presence of an
electron, the conduction band would be stabilized owing to
fluctuations of the lattice's polarization, much like that leading to
the attractive van der Waals force. In a rather distinct treatment of
non-polar substances, the latter author assumes that optical phonons,
if any, are not even important and subsequently estimates the gap
narrowing exclusively due to interaction between electrons and
acoustic phonons. Hereby, the issue of the surprisingly large value of
$\gamma$ is thus relegated to the detailed question of electron
scattering off lattice vibrations.

Early efforts to explain the Urbach tail {\em also} emphasized the
importance of lattice polarization, such as that arising from
excitonic effects. Subsequent studies however seem to shift their
focus to the physics of Anderson localization, by looking at the
electronic motion subject to a random potential. Multiple studies
spanning decades indicate that the ``simple'' exponential shape of the
optical edge is nothing but: Within a general, even if
phenomenological picture, the exponential tail can be viewed as an
intermediate regime between weak and very strong localization of
electrons subject to a random potential. To account for the
substantial spectral range occupied by the Urbach tail, one must
however assume that the spatial correlations between the fluctuations
of the random potential must be strong within a small multiple of the
atomic size but decay rapidly beyond that, more rapidly than, for
instance, exponentially.  The microscopic mechanism behind this
puzzling observation is not clear. Still, relatively explicit
description modeling polaronic effects indicate that the exponential
density of localized states is in fact robust, subject to modest
non-adiabatic effects.~\cite{PhysRevB.41.7641}

\begin{table}
\begin{tabular}{|l|c|c|c|c|c|}
  \hline 
  ¥ & $\gamma$, eV/K & $e^{\gamma/2}$&$\alpha$, cm$^{-1}$ & $\omega$, K & $V$, eV\\ 
  \hline 
  $a$-SiO$_2$ & $1.15\cdot 10^{-3}$ & 785&$2.40\cdot 10^{2}$ & 674 & 11.6\\ 
  \hline 
  $c$-SiO$_2$ &  $0.98\cdot 10^{-3}$ & 303&$4.70\cdot 10^{3}$ & 535& 10.7\\ 
  \hline 
  $a$-As$_2$S$_3$ &  $0.78\cdot 10^{-3}$ & 93 & $2.35\cdot 10^{2}$ & 275 & 4.8 \\ 
  \hline 
  $a$-As$_2$Se$_3$ & $1.41\cdot 10^{-3}$ & 3735  & N/A&N/A & 6.3\\ 
  \hline 
\end{tabular} 
\caption{\label{VTable} The quantity $\gamma$ is from Eq.~(\ref{EgTlinear}). 
  The quantity $e^{\gamma/2}$ is computed using the dimensionless $\gamma$. 
  The quantity $\alpha$ gives the value of the absorption coefficient, at which
  the optical gap was determined, see Eq.~(\ref{alpharule}). Values
  $T_{\rm m}$ used for the purpose of fitting are, respectively, 1986~K for 
  SiO$_{2}$~\cite{haynes2011crc},
  585~K for As$_{2}$S$_{3}$~\cite{haynes2011crc,
    doi:10.1002/pssb.2220640116} and 625~K for 
  As$_{2}$Se$_{3}$~\cite{ARAI1975295}.}
\end{table}

Here we present a qualitative microscopic picture that directly
connects the phenomena of both band narrowing and the tail of
localized states with the atomic motions. In turn, these are tied to
the fluctuations of the optical phonons. The effective random
potential is seen as stemming directly from the disorder in the bond
length. The coupling with vibrations comes about because both the
electronic transfer integral and the dipole moment for electronic
motions depends on the bond length.

We elucidate the apparent, seemingly intrinsic connection between the
presence of an insulating gap, on the one hand, and the existence of
optical phonons, on the other hand. The physical reason behind this
connection is relatively lucid: In the presence of bonding that is not
closed-shell, electronic excitations across the forbidden gap and
polarization must both involve charge transfer between
atoms. Accordingly, the existence of an insulating gap or pseudo-gap
necessarily implies that there are at least two orbitals centered on
distinct atoms within the primitive cell and so the optical phonons
must be present, too.  A corollary of this observation is that if the
optical phonons are absent, the solid is metallic. Hence an elemental
solid that has one atom per primitive cell will be ordinarily a
metal. An arguable exception to this rule would be when such an
elemental solid hosts a charge density wave; still such a solid,
strictly speaking, should be regarded as hosting at least two
inequivalent atoms per primitive cell and would, in fact, possess
optical phonons. These notions are consistent with
experience. Applying the present argument to the Wigner crystal leads
one to conclude that the ground state of this quantum solid must be
anti-ferromagnetic, consistent with quantum Monte Carlo
simulations.~\cite{PhysRevB.70.094413, PhysRevLett.102.126402}

As an additional dividend, the present picture allows one to build a
field theory, for a substantial class of materials, to show that the
optical phonons can be thought of as fluctuations of the order
parameter for the metal-insulator transition.  The order parameter
must be at least two-dimensional, the two components reflecting,
respectively, the strength of spatial inhomogeneity of the charge
distribution on the atomic sites and inter-atomic spacings. (The
inhomogeneity can be thought of spatially-inhomogeneous bond
saturation.) As such, the gap narrowing can be viewed rather generally
as signaling the eventual transition to a metallic state. In the
present treatment, the metal-to-insulator transition is naturally a
symmetry breaking transition and could be either continuous or
discontinuous. Despite the order parameter being vectorial and
possibly exhibiting a continuous symmetry in the symmetric state, the
symmetry breaking can be explicitly shown to not give rise to
Goldstone modes. Instead, the optical phonons, which emerge following
the symmetry breaking are gapped excitations as they should be. 


\section{The $T$-dependence of the optical gap and the gap edge, and
  the identity of the charge transfer states}
\label{empirics}

The main purpose of this Section is to substantiate our earlier
statement is that the $T$-dependence of the band gap is surprisingly
strong, with implications for calculations of the electrical
conductivity. The dependence of the {\em optical} gap can be described
by a simple phenomenological relation:~\cite{PhysRevLett.47.1480,
  PhysRev.78.808.2, PhysRev.82.900, doi:10.1063/1.104723}
\begin{equation} \label{EgTcoth} E_g = C_1 - (\gamma/T_0) \coth(T_0/T)
\end{equation}
where $C_1$ and $T_0$ are numerical constants. At sufficiently high
temperatures, this implies a simpler yet, linear $T$-dependence in
Eq.~(\ref{EgTlinear}). Functional forms other than that in
Eq.~(\ref{EgTcoth}) often work, too.~\cite{VARSHNI1967149,
  TICHY1992198} As mentioned in Section~\ref{motivation}, one
customarily determines the optical gap at a fixed value of the
absorption coefficient $\alpha$ because the optical edge is not
sharp. The coefficients $C_1$ and $C_2$ thus depend on the precise
value of $\alpha$.

In a relatively broad energy range---often corresponding to several
orders of magnitude in terms of $\alpha$---the optical edge apparently
exhibits an approximately exponential shape:
\begin{equation} \label{alphaE} \alpha(E) =  \alpha_0 e^{(E-E_0)/\epsilon_\tU}
\end{equation}
where the quantity $\epsilon_\tU$ reflects the width of absorption
edge. The temperature dependence of the width of the Urbach tail is
often well described by a phenomenological
relation:~\cite{doi:10.1002/pssb.2221160133}
\begin{equation} \label{eU} \epsilon_\tU = C \omega \coth(\omega/T)
\end{equation}
where the constant $C$ is numerically close to
two.~\cite{doi:10.1002/pssb.2221160133}

Whether or not the substance obeys the noteworthy trend illustrated in
Fig.~\ref{alphaCross}, it is clear from Eqs.~(\ref{EgTcoth}),
(\ref{alphaE}) and (\ref{eU}) that the value of $\gamma$ generally
depends on the value of the absorption coefficient at which the
optical gap is defined. One may further ask: What is the value of the
band gap that is relevant in the context of electrical conduction or
whether such a question is well-posed in the first place.

To avoid ambiguity we note that, strictly speaking, the exponential
shape of the absorption edge is not equivalent to the edges of the
bands themselves being exponential. Still, photo-conductivity
experiments by Monroe and Kastner~\cite{PhysRevB.33.8881} strongly
suggest that the individual bands do have several decades worth of
density of states in the form of exponential tails. It is
straightforward to show that the width of the exponential tail in the
optical absorption is equal to the sum of the widths $\epsilon_{c}$
and $\epsilon_{v}$ of the exponential edges of the conduction and
valence bands, respectively:
\begin{equation} \label{eUcv} \epsilon_\tU = \epsilon_{c} +
  \epsilon_{v}.
\end{equation}

We now briefly discuss two specific views on charge transport in
semiconductors.~\cite{Emin_rev, Emin_revII} In one model, the
conduction is solely due to delocalized carriers. Hence the energy
scale $E_\sigma$ in the activated part of the electrical conductivity,
$e^{-E_\sigma/k_B T}$, is determined by the energy difference $(E_c -
\mu)$ between the bottom of the conduction (mobility) band and the
chemical potential $\mu$, for the electrons, and the energy difference
$(\mu - E_v)$ between the chemical potential and the ceiling of the
valence (mobility) band, for the holes.

In an alternative picture, charge transport is carried out via
activated hopping of small polarons. These quasi-particles result from
self-trapping of electrons,~\cite{PhysRevLett.36.323} which requires
sufficiently strong electron-phonon coupling and would be enhanced by
the presence of disorder.~\cite{PhysRevB.49.14290, ZhangPhillips} It
is deemed that the materials of the type exemplified in
Fig.~\ref{EgTFig} do in fact conduct electricity primarily with small
polarons, but there does not seem to be complete agreement on this
issue, see Refs.~\onlinecite{Emin_rev, Emin_revII, Emin_Hall} for a
detailed discussion.

In the most basic version of the small polaron picture, the activation
part of the $T$-dependence of the electrical conductivity due to, say,
the electron polarons is determined by energy cost of putting an
electron in the small polaron band, $(E_{e, \text{pol}} - \mu)$, and
the activation energy needed to reach the pertinent mobility band from
the polaron state: $(E_c - E_{e, \text{pol}})$. As in the
delocalized-carrier picture, the activation energy is $(E_c - \mu)$.
In the presence of a ``smeared'' edge in the density of states, this
relatively simple picture deserves some additional discussion. Indeed,
the activation exponent for transport must be averaged with respect to
the DOS of the localized states that could be transition states for
polaron hopping. Considering electronic polarons for concreteness, one
may write
\begin{equation} \label{intE} \la \nu e^{-(E - E_{e, \text{pol}})/k_B
    T} \ra \propto \int \!\! dE d^3 \br \, \nu \, e^{-E/k_B T} n(E,
  \br),
\end{equation}
where the coordinate dependence in the density of states $ n(E, \br)$
reflects that the wave-function overlap between the initial and
transition states will generally depend on spatial separation between
the two.  The specifics of averaging depend on the detailed nature of
the disorder. Two limiting cases are relatively straightforward to
describe:

(1) If the disorder is caused exclusively by vibrations of the
lattice, then the effective density of states $n(E, \br)$ becomes
coordinate-independent.  This is because the waiting time for polaron
hopping is much longer than the time scale of the lattice vibrations.
Thus every configuration of the lattice will be sampled locally so
that the volume-integrated density of states is proportional to the
{\em bulk} density of states $e^{E/\epsilon_{c}}$ within the tail:
\begin{equation} \label{nEdynamic} \int d^3\br \, n_\text{dynamic}(E,
  \br) \propto e^{(E-E_c)/\epsilon_c},
\end{equation}
where $E < E_c$. Note that the prefactor $\nu$ for the rate of hopping
is largely determined by the vibrational frequency: The electronic
wave function overlap is large implying the hopping is adiabatic.

Now, according to the discussion of Eq.~(\ref{alphaE}), $\epsilon_{c}$
is only slightly greater, if at all, than the temperature. This
implies that although the integral in Eq.~(\ref{intE}) is dominated by
the states near $E_c$, a broad spectral range of states will generally
contribute to transport as transition states for polaron hopping. In
any event, insofar as $\epsilon_{c} < k_B T$, the overall activation
energy scale for the polaron hopping is still determined by the
position of the pertinent mobility band since the integrand in
Eq.~(\ref{nEdynamic}) peaks out near $E_c$: The density of states in
the mobility band, i.e. at $E > E_c$, depends on energy much more
slowly than the Boltzmann weight $e^{-E/k_B T}$.

(2) When the disorder is exclusively static, the coordinate-averaged
effective density of transition states decreases away from the band
edge, with decreasing energy, than the exponential density of states
in the r.h.s. of Eq.~(\ref{nEdynamic}). This is because a tail state
at a given energy is no longer guaranteed to be near the initial
location of the polaron and so the probability rapidly decreases with
the depth $(E_c - E)$ of the state in question. Indeed, the electronic
wave-function overlap between the initial and transition states will
go (for electrons) as $e^{- r/l}$, where $l$ is the localization
length of the orbital representing the transition state. Thus the
spatial averaging in Eq.~(\ref{intE}) yields a factor $e^{-
  \bar{r}/l}$, where the typical distance $\bar{r}$ to the nearest
transition state at energy $E$ is determined from the bulk density of
states from the r.h.s. of Eq.~(\ref{nEdynamic}) according to
$1/\bar{r}^3 \sim l^{-3} e^{(E-E_c)/\epsilon_c}$. This yields
\begin{equation} \label{nEstatic} \int d^3 \br \, n_\text{static}(E,
  \br) \propto \exp[-e^{(E_c - E)/3\epsilon_c}],
\end{equation}
where $E < E_c$, as before.  The r.h.s. is only a (rough) upper bound,
and increasingly so for smaller values of $E$. This is because for
larger values of $r$, and hence large values of $(E_c - E)$, the
transition will be increasingly more non-adiabatic as the transition
state wave-function becomes more and more localized. Consequently, the
frequency prefactor for the hopping rate $\nu$ will be decreased by
the corresponding Franck-Condon factor for the lattice
vibrations. Thus the dependence of the coordinate-integrated density
of transition states on $E$ on energy is even stronger than that
implied by the r.h.s. of Eq.~(\ref{nEstatic}). Because of this very
strong dependence, the lower limit for the energy integration in
Eq.~(\ref{intE}) is now effectively pinned at or just below the
mobility edge. The coordinate-averaged density of states for the
transition states for polaron hopping in amorphous materials will
interpolate between Eqs.~(\ref{nEdynamic}) and (\ref{nEstatic}), the
detailed answer depending on the interplay of the static and dynamic
disorder.

In any event, we conclude that the activation energy for the
electrical conductivity is determined by the locations of the mobility
edges $E_c$ and $E_v$. Consequently, one may peg the relevant value of
the absorption strength at the corresponding energy difference:
\begin{equation} E_g = E_c - E_v.
\end{equation}
The spacing between the mobility edges is conventionally determined in
absorption experiments by first fitting the large-$\alpha$ portions of
the absorption spectra by a quadratic functional form $(E -
\omega_b)^2$ where $\omega_b$ is a fitting parameter. Absorption
strength at $\omega_b$ is often numerically close to $10^{3}
\text{cm}^{-1}$.  The aforementioned quadratic form is
expected~\cite{MottDavis1979, Tauc1974} when the conduction and
valence bands, respectively, behave like those in a periodic 3D solid:
$\propto (E - E_c)^{1/2}$ and $\propto (E - E_v)^{1/2}$. We however
believe that this empirical procedure overestimates the band gap
because the states in the mobility bands are stabilized owing to
electron-lattice interactions as discussed in
Section~\ref{motivation}, the amount of stabilization increasing
toward the band edge. (This immediately follows from the expression
for the second-order term in the perturbation
theory.~\cite{LLquantum}) This will lead to deviations from the simple
square-root shape of the band edge expected of a strictly periodic
solid in which nuclei do not vibrate altogether.

Here, instead, we determine the spacing between the mobility edges by
utilizing the results of S.~John et al.~\cite{PhysRevLett.57.1777,
  PhysRevB.37.6963, 5390028} obtained for a one-electron model, in
which the electron is subject to a random potential $V(x)$:
\begin{equation} \label{Hrandom} [-\hbar^2/2m + V(x)] \psi(x) = E
  \psi(x).
\end{equation}
By construction, $\la V(x) \ra = 0$, while the corresponding
auto-correlation function decays in a Gaussian fashion with distance:
\begin{equation} \label{CFGaussian} \la V(x) V(0) \ra = V_\text{rms}^2
  e^{-x^2/L^2}.
\end{equation}
The shape of the tail of the localized shape depends on the detailed
functional form of the auto-correlation. A substantial exponential
tail is predicted for functional forms that decay faster than the
Gaussian, but not slower.  With this proviso, the model yields both
the stabilization $\Delta E$ of the band edges and the width
$\epsilon$ of the exponential tail, the two being exactly proportional
to each other with a universal coefficient that depends on the
dimensionality of space, very much consistent with the behavior seen
in Fig.~\ref{alphaCross}. According to Eqs.~(A4) of
Ref.~\onlinecite{PhysRevB.37.6963},
\begin{equation} \label{stab} \Delta E = - V_\text{rms}^2/2
  \epsilon_L,
\end{equation}
while Eq.~(6) of Ref.~\onlinecite{PhysRevLett.57.1777}
states that 
\begin{equation} \label{tail} \epsilon = V_\text{rms}^2/14.4
  \epsilon_L.
\end{equation}
Here the quantity $\epsilon_L$ reflects the cost of localizing the
particle within a region of size $L$:
\begin{equation} \epsilon_L \equiv \hbar^2/2mL^2.
\end{equation}
In conjunction with Eq.~(\ref{alphaE}), this immediately yields a
simple relation:
\begin{equation} \label{alpharule} \epsilon_\tU = (E_0 - E_g)/7.2.
\end{equation}

Notwithstanding the approximate nature of the calculation in
Refs.~\onlinecite{PhysRevLett.57.1777, PhysRevB.37.6963, 5390028}, the
quantities $\epsilon_\tU$ and $(E_0 - E_g)$ are obtained
internally-consistently using a constructive argument. Thus the above
{\em relation} between the two quantities is expected to be less
sensitive to the approximations made in
Refs.~\onlinecite{PhysRevLett.57.1777, PhysRevB.37.6963, 5390028} than
the individual quantities themselves. In any event---and consistent
with expectation---the so determined gap is lower than than obtained
using the purely phenomenological protocol described above, i.e., at
values of $\alpha$ close $10^2~\text{cm}^{-1}$. This yields rather
substantial values for the coefficient $\gamma$ from
Eq.~(\ref{EgTlinear}). As discussed in Section~\ref{motivation}, this
has significant consequences for quantitative estimates of electrical
conductivity in semiconductors.

\section{Band narrowing as a result of fluctuations of optical
  vibrations: A qualitative treatment}
\label{optical}

Next we outline a simple structural model for the coupling of the band
gap to lattice distortion.  Our main focus is on two relatively
distinct types on insulators: (1) Non-polar or weakly polar
semiconductors exhibiting spatial inhomogeneous bond saturation, such
as the chalcogenide compounds As$_2$Se$_3$ and As$_2$S$_3$, see
Fig.~\ref{EgTFig}. In such compounds, there are two types of
interactions: covalent bonds and the secondary
interactions~\cite{Pyykko} that are stronger than Van der Walls but,
technically, are closed-shell. The total number of each is roughly the
same.~\cite{ZLMicro1} (2)~Ionic insulators such as oxides or halides,
such as SiO$_2$, see Fig.~\ref{EgTFig}. We will also briefly touch on
yet another class of compounds. These are non-polar or weakly polar
semiconductors that exhibit largely uniform bond strength throughout,
such as Si, Ge, or GaAs. All individual bonds in the latter compounds
are well described as canonical covalent bonds of order 1.

Type 1 compounds are well represented by elemental arsenic and various
chalcogenide alloys: These can be thought of approximately as
$pp\sigma$-bonded networks exhibiting distorted octahedral
coordination.~\cite{ZLMicro1} In turn, these distorted structures can
be thought of as symmetry lowered versions of parent structures
defined on the simple cubic lattice, possibly with vacancies. In the
absence of $sp$-mixing, one can think of the individual chains
comprising the cubic lattice as Peierls unstable toward
dimerization.~\cite{ZLMicro1, burdett5764, 0022-3719-13-26-010} The
presence of $sp$-mixing does affect the eventual geometry of the
distorted lattice but does not change the picture qualitatively. This
is consistent with the $s$- and $p$-bands being well separated in the
electronic density of states for these compounds, the $s$-band being
completely filled, see for instance Ref.~\onlinecite{LL1}.

Qualitatively, the electronic transfer integral scales exponentially
with the interatomic distance
\begin{equation} \label{tr} t_\text{e} \approx t_0 e^{-A r},
\end{equation}
for an interatomic distance $r$ that is not too small and $A$ is on
the order of the inverse Bohr radius; the quantity $A r$ exceeds one
but is certainly less than 10.  The value of $t$ is of the order
eV. To give a specific example, $A \approx 1.2$~\AA$^{-1}$ for the
As-Se bond, as can be inferred~\cite{GHL} from the semi-empirical
package MOPAC.~\cite{mopac, mopacPM7} In the amorphous As$_2$Se$_3$
compound, the stronger and weaker bond are typically of length $r_1 =
2.4$~\AA and $r_2 = 3.6$~\AA, respectively.~\cite{LL1} Direct sampling
of the amorphous As$_2$Se$_3$ samples generated in Ref.~\cite{LL2}
indicates the corresponding $pp\sigma$ transfer integrals are
typically 6~eV and 1-2~eV for the strong and weak As-Se bond,
respectively.

Clearly, fluctuations in the inter-atomic spacings, due lattice
vibrations, will modulate the magnitude of the transfer matrix
elements from Eq.~(\ref{tr}). In the Born-Oppenheimer approximation,
electrons will be subject to a static random potential such as that in
Eq.~(\ref{Hrandom}). To find this connection, we consider for
concreteness a $pp\sigma$-bonded solid made of 1D chains at
half-filling. Each atom houses three $p$ orbitals; the chains along
the $x$ axis are made of $p_x$ orbitals etc. It will suffice for now
to assume that there is no $sp$-mixing, so that the chains are
non-interacting. For each chain, we write down a H\"uckel energy
function, one orbital per site:
\begin{eqnarray} \cH = \sum_n \sum_{s=\pm1/2} \hspace{0mm} &\left[ -
    t_{n, n+1} \, (c^\dagger_{n,s} c_{n+1,s}^{} + c^\dagger_{n+1,s}
    c_{n,s}^{}) \right. \nonumber \\ &+ \left. \varepsilon_n \,
    c_{n,s}^\dagger c_{n,s}^{} \right],
  \label{Huckel}
\end{eqnarray}
where $c_{n, s}^\dagger$ ($c_{n, s}^{}$) creates (annihilates) an
electron on site $n$ with spin $s$.  We will assume that the electrons
are non-interacting. Within limits, such interactions can be thought
of as renormalizing the transfer integrals $t_{n, n+1}$ and the
on-site energies $\varepsilon_n$.~\cite{ZLMicro1} The transfer
integrals are assumed to scale with the interatomic spacing as in
Eq.~(\ref{tr}).

The chain described by Eq.~(\ref{Huckel}), at half-filling, is
unstable toward Peierls dimerization,~\cite{Peierls,
  RevModPhys.60.781, ABW} if the electronegativity variation, as
embodied in the on-site energies $\varepsilon_n$, is not too
large.~\cite{PhysRevLett.49.1455} Such an insulator can be said to
host a ``bond-order'' charge density wave. The dimerized chain
contains two atoms per repeat unit and, also, supports optical
phonons. It is convenient to introduce a specific modulation pattern
for the electronegativity, which has the same periodicity as the
dimerized chain:
\begin{equation} \label{epsn} \varepsilon_n = (-1)^n \varepsilon.
\end{equation}
If desired, one may create a primitive unit that spans more than two
atoms, by employing a different value for the electron filling and the
corresponding period of the electronegativity variation.

\begin{figure}[t!]
\center
\includegraphics[width = .85 \figurewidth]{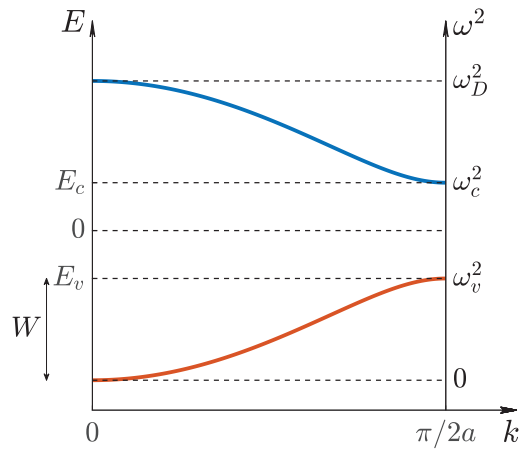}
\caption{\label{spectrumSketch} The electronic and vibrational
  spectrum of a dimerized chain. The dimerization could be due to an
  alternation in the strength and/or identity of the ion.}
\end{figure}

We reiterate that for a sufficiently small $\varepsilon$, we have an
insulator of type 1. Otherwise, i.e., when $|\varepsilon| \gtrsim
\bar{t}$,~\cite{PhysRevLett.49.1455} the chain exemplifies an
insulator of type 2. In a dimerized chain, the hopping matrix element
alternates between two values:
\begin{equation} \label{tnn}
  t_{n, n+1} \equiv \bar{t} + (-1)^n t, \hspace{3mm} |t| \le \bar{t}.
\end{equation}
In the existing notation, $t = (t_1 - t_2)/2$ and $\bar{t} = (t_1 +
t_2)/2$. Hamiltonian (\ref{Huckel}) can be broken up, in a standard
fashion, into a set of $N/2$ non-interacting two-state systems. Each
such system operates on two plane waves, each at wavevector $k$,
spanning the odd- and even-numbered sites, respectively:~\cite{ABW}
\begin{align} \label{odd} | \psi_k^\text{odd} \rangle &= (2/N)^{1/2}
  \sum_n | \psi_{2n+1}^{} \rangle e^{ika \, (2n+1)} \\ \label{even} |
  \psi_k^\text{even} \rangle &= (2/N)^{1/2} \sum_n | \psi_{2n}^{}
  \rangle e^{ika \, 2n}.
\end{align}
The kets $| \psi_n \rangle$ stand for the basis set of the matrix
$H_{mn}^{}$ defining the tight-binding Hamiltonian from
Eq.~(\ref{Huckel}): $\cH \equiv \sum_{mn, s} H_{mn}^{} c_{m,
  s}^\dagger c_{n, s}^{}$. The two-level system, for each value of
$k$, is defined by the matrix:
\begin{equation} \label{TLS} \left( \begin{array}{c|c} - \varepsilon & -2
      \bar{t} \cos ka + 2 i t \sin ka \\ \hline -2 \bar{t} \cos ka - 2
      i t \sin ka & \varepsilon
    \end{array}
  \right),
\end{equation}
where $a$ is the lattice spacing.  The resulting spectrum is:
\begin{equation} \label{spectrumEq} E_k = \mp [\varepsilon^2 + 4t^2 +
  4(\bar{t}\,^2 - t^2) \cos^2ka]^{1/2},
\end{equation}
see the sketch in Fig.~\ref{spectrumSketch}. The spectrum is doubly
degenerate: $E_{k} = E_{-k}$ and possesses a band gap:
\begin{equation} \label{EgChain} E_g = 2 \sqrt{(t_1 - t_2)^2 +
    \epsilon^2}.
\end{equation}

The Hamiltonian (\ref{Huckel}), which is defined for fixed values of
$\bar{t}$, $t$, and $\varepsilon$, can be generalized for parameters
$t(x)$ and $\epsilon(x)$ slowly varying with the spatial coordinate
$x$ along the chain.~\cite{PhysRevB.21.2388, PhysRevLett.49.1455,
  ZL_JCP} This is conveniently done by shifting the wavevector
reference in the two-level Hamiltonian (\ref{TLS}) to the edge of the
Brillouin zone of the original Hamiltonian (\ref{Huckel}), $(k -
\pi/2a) \to k$, and then expanding around $k = 0$:
\begin{equation} \label{Hcont} \cH = - i 2 \bar{t} a \, \hat{\sigma}_3
  \frac{\prtl}{\prtl x} - 2 t(x) \, \hat{\sigma}_1 - \varepsilon(x) \,
  \hat{\sigma}_2,
\end{equation}
where the operators $\hat{\sigma}_i$ are the standard Pauli
matrices. We have relabeled the latter, $(\sigma_1, \sigma_2,
\sigma_3) \to (\sigma_2, \sigma_3, \sigma_1)$, so that Hamiltonian
(\ref{Hcont}) now operates on spinors with components $\psi_1 =
(\psi^\text{odd} + \psi^\text{even})/\sqrt{2}$ and $\psi_1 =
(\psi^\text{odd} - \psi^\text{even})/\sqrt{2}$. These two correspond
with the filled and empty orbitals of the non-dimerized chain and
sometimes are called the right and left movers. This is because of the
aforementioned degeneracy $E_{k} = E_{-k}$, and the equivalence of
points $k$ and $k+\pi/a$. 

It is convenient to rotate the basis set in Eq.~(\ref{Hcont}) in the
$(\sigma_1, \sigma_2)$ plane so that the corresponding part of the
Hamiltonian is directed along, say, direction $\sigma_1$. Keeping in
mind that the rotation angle $\varphi$ is $x$-dependent, one obtains:
\begin{equation} \label{Hcont2} \cH = - i 2 \bar{t} a \,
  \hat{\sigma}_3 \frac{\prtl}{\prtl x} - \Delta \, \hat{\sigma}_1 -
  \bar{t} a \frac{\prtl \varphi}{\prtl x},
\end{equation}
where
\begin{equation} \label{DeltaDef} \Delta \equiv [(2 t)^2 +
  \varepsilon^2]^{1/2}
\end{equation}
and
\begin{equation} \frac{\prtl \varphi}{\prtl x} = \frac{2}{\Delta^2}
  \left(\varepsilon \frac{\prtl t}{\prtl x} -t \frac{\prtl
      \varepsilon}{\prtl x} \right).
\end{equation}
The Schr\"odinger equation for the Hamiltonian (\ref{Hcont2}), which
operates on the spinor with components $\psi_1$ and $\psi_2$, can be
profitably recast in terms of the linear combinations $\psi_\pm =
(\psi_1 \pm i \psi_2)/\sqrt{2}$:
\begin{align} \label{pm} i \left( - 2 \bar{t}a \frac{\prtl}{\prtl x} +
    \Delta \right) \psi_- &= \left(E + \bar{t} a \frac{\prtl
      \varphi}{\prtl x} \right) \psi_+ \\ i\left( - 2 \bar{t}a \,
    \frac{\prtl}{\prtl x} - \Delta \right) \psi_+ &= \left(E + \bar{t}
    a \frac{\prtl \varphi}{\prtl x} \right) \psi_-,
\end{align}
which then allows one to straightforwardly generate a second-order
differential equation for either $\psi$.

It is instructive to consider first the relatively simple case of a
purely ionic ordered compound exhibiting no dimerization, so that $t =
\prtl t/\prtl x = 0$, while $\bar{t} = \text{const}$.  In this case,
one obtains a Klein-Gordon equation for $\psi_+$:
\begin{equation} \label{KGionic} \left( - 4\bar{t}^2 \frac{a^2
      \prtl^2}{\prtl x^2} + \varepsilon^2 - 2 \bar{t} a \frac{\prtl
      \varepsilon}{\prtl x}\right) \psi_+ = E^2 \psi_+,
\end{equation}
and an analogous equation for $\psi_-$, not shown.

Generally, however, one obtains a more complicated equation, which can
be interpreted as a Klein-Gordon equation with the energy reference
point shifted off the classical vacuum state:
\begin{equation} \label{KG} \left( - 4\bar{t}^2 \frac{a^2
      \prtl^2}{\prtl x^2} + \Delta^2 - 2 \bar{t} a \frac{\prtl
      \Delta}{\prtl x} \right) \psi_+ = \left(E + \bar{t} a
    \frac{\prtl \varphi}{\prtl x} \right)^2 \psi_+
\end{equation}
and an analogous equation for $\psi_-$, not given.

In going from Eq.~(\ref{pm}) to Eq.~(\ref{KG}), we omitted the
following contribution to the l.h.s.:
\begin{equation} \label{omitted} - 2 a^2 \frac{\prtl \bar{t}^2}{\prtl
    x} \, \frac{\prtl \psi_+}{\prtl x} + \frac{2 \bar{t} a (\bar{t}a
    +\Delta \psi_+) }{(E + \bar{t} a \prtl \varphi/\prtl x)^2}
  \frac{\prtl}{\prtl x}\left(\bar{t} \frac{\prtl \varphi}{\prtl x}
  \right).
\end{equation}
The first term will contribute in the forth order in $k^4$ because the
quantity $(\prtl \bar{t}^2/\prtl x)$ has a random sign. This
circumstance will also automatically take care of the term itself
being anti-hermitian. A contribution of order $k^4$ would exceed the
accuracy of the long-wavelength approximation that led to
Eq.~(\ref{Hcont}) in the first place, hence we have omitted it.  The
second term can be analyzed analogously.  For these same reasons,
importantly, the quantity $\bar{t}^2$ in front of the second
derivative in Eq.~(\ref{KG}) should be regarded as a constant (though
not necessarily equal to its average value!).

Apart from the ``shift'' term in the square brackets, Eq.~(\ref{KG})
is well known,~\cite{PhysRevB.21.2388, PhysRevLett.49.1455} of course,
as part of continuum descriptions of topological midgap states in
conjugated polymers~\cite{RevModPhys.60.781} and midgap states in
amorphous chalcogenide alloys.~\cite{ZL_JCP, ZLMicro2} In those
descriptions, the electronegativity variation was considered as
built-in and constant, as pertaining to a set of ions are rigidly
assigned to the lattice. Here, we consider $\varepsilon$ as spatially
varying to include the possibility of local polarization
fluctuations. We observe that upon including this possibility, our
``Klein-Gordon'' equation is no longer an eigen-value problem. In
fact, one anticipates that for sufficiently large values of
fluctuations, the ``electron'' and ``hole'' (or ``positron'')
solutions of Eq.~(\ref{KG}) will mix implying an insulator-to-metal
transition, owing to polarization fluctuations. This is consistent
with the mechanism of the emergence of excitonic
insulator~\cite{PhysRevLett.19.439} states and is an encouraging sign
indeed.

Equations analogous to Eqs.~(\ref{KG}) and (\ref{KGionic}) can be
written for the spatial coordinates $y$ and $z$ just as well. Added
together, these equations yield an analogous equation to
Eq.~(\ref{Hrandom}). Indeed, for small deviation of the energy from
the (stabilized) bottom of the respective mobility band, call it
$E_c$, Eq.~(\ref{KG}) becomes linear in the energy $E$. Upon a change
of variables $E \to E + E_c$, leaving in only the linear term in $E$,
and noting that $E_c \sim \Delta$, we obtain that the magnitude of the
fluctuation of the random potential goes with the quantity $\Delta/2$
plus the (numerically comparable) contributions of the $\bar{t} a
\prtl \Delta/\prtl x$ term in Eq.~(\ref{KG}). These quantities are
straightforwardly related to the local variation of the bond strength
and electronegativity, c.f. Eqs.~(\ref{EgChain}) and
(\ref{DeltaDef}). Inasmuch as the distributed potential in
Eqs.~(\ref{KG}) and (\ref{KGionic}) change approximately linearly with
the inter-atomic distances within the appropriate range, $V_\text{rms}
\propto \delta r$, their fluctuations are approximately Gaussian,
too. Thus we can carry over the predictions (\ref{stab}) and
(\ref{tail}), due to Refs.~\onlinecite{PhysRevLett.57.1777,
  PhysRevB.37.6963, 5390028}, to the present context. Furthermore, the
amplitude of those fluctuations should be proportional to that of a
vibrating bond:
\begin{equation} \label{Vr} V_\text{rmp} \sim V \frac{\delta r}{a},
\end{equation}
where $\delta r$ is the typical vibrational amplitude:
\begin{equation} \frac{\delta r}{a} \le \left. \frac{\delta r}{a}
  \right|_{T = T_m} \simeq 0.1.
\end{equation}
The latter approximate equality expresses the Lindemann criterion of
melting, by which the relative vibrational displacement is universally
about one tenth near the melting temperature $T_m$.~\cite{L_Lindemann,
  PinesEXS} The energy parameter $V$ corresponds to the variation of
the (local) band gap with bond deformation corresponding to long
wavelength optical phonons, as just discussed:
\begin{equation} \label{Vlogr} V \simeq \frac{\Delta}{\prtl \ln r},
\end{equation}
where $r$ is the bond length, as in Eq.~(\ref{tr}).  We anticipate
that $V$ is numerically close to several eV. Indeed, according to
Eq.~(\ref{DeltaDef}) and the discussion following Eq.~(\ref{tr}),
fluctuations in the quantities $t_1$ and $t_2$ should contribute
together a few 1 eV for chalcogenides. We expect a comparable
contribution from fluctuations in $\epsilon$. There will be some
additional enhancement because of space dimensionality is three;
Eq.~(\ref{DeltaDef}) is for a 1D chain.

In view of the standard result~\cite{LLstat}
\begin{equation}\label{Eq:deltaR} \delta r^{2} =
  \frac{\hbar}{2m\omega} \coth\left( \frac{\hbar\omega}{2
      k_{B}T}\right),
\end{equation}
relation (\ref{Vr}) justifies the functional forms in
Eqs.~(\ref{EgTcoth}) and (\ref{eU}) used for the temperature
dependences of the band gap and the width of the Urbach tail. We show
the corresponding fits of the band gap for the substances shown in
Fig.~\ref{EgTFig}, alongside the experimental data. For the purpose of
fitting, we have adopted the size of the rigid molecular unit---often
called the ``bead''---for the length $L$. By
construction,~\cite{LW_soft} motions on lengthscales greater than the
bead size, the liquid behaves as that composed of weakly interacting
particles, such as in Lennard-Jones systems. This is very much
consistent with the definition of $L$ in Eq.~(\ref{CFGaussian}).
Ordinarily, the bead contains several atoms. We took the bead size for
quartz from Ref.~\cite{LW_soft}: $L=3.6$~\AA; the bead essentially
corresponds with the SiO$_{4/2}$ unit. Likewise for the chalcogenides,
we used the argument from Ref.~\cite{ZLMicro2} that the bead
corresponds to the AsS$_{3/2}$ ($L=3.9$~\AA) and AsSe$_{3/2}$
($L=4.1$~\AA) pyramids. Explicit knowledge of the mass $m$ from
Eq.~(\ref{Eq:deltaR}) is not needed since the value of $\delta r/a$ is
calibrated so that it achieves exactly $1/10$ at the melting
temperature $T_m$. The latter values are listed in the caption of
Table~\ref{VTable}.

The resulting values of the fitting parameters $V$ and $\omega$ from
Eq.~(\ref{Eq:deltaR}) are provided in Table~\ref{VTable}; they are
very much in accord with expectation. The parameter $V$ for quartz is
large than for the chalcogenides, which is consistent with its larger
band gap.

The mapping between the present model and that from
Refs.~\onlinecite{PhysRevLett.57.1777, PhysRevB.37.6963, 5390028} is
far from exact. The auto-correlation of the noisy potential depend on
the Cartesian distances between distinct locations.  Yet for
non-interacting chains, such correlations will be found only for pairs
of points located on the same chain. Still, the chains must interact
because of $sp$-mixing (or its analogs, if other types of orbitals are
involved), various interactions among electrons and ions, and the for
the simple reason that the lengths of distinct bonds emanating from an
atom change all at the same time as the atom moves about. As a result,
the random potential in the 3D version of Eqs.~(\ref{KG}) and
(\ref{KGionic}) will no longer be separable into functions of
individual coordinates $x$, $y$, and $z$. Instead, the
auto-correlation function for the fluctuations will be depend on the
Cartesian distances between distinct locations, at least for not too
long distances. In any event, we do not expect the predictions of the
random-potential model to depend much on the precise spatial shape of
the potential well of size $L$ so long as the well itself has
sufficiently steep walls, see discussions in
Refs.~\onlinecite{PhysRevB.37.6963, 5390028}.

We observe that at this level of approximation, the fluctuations of
the random potential are coupled to lattice motions that modify the
differentials in the transfer integral and electronegativity largely
{\em within} the unit cell, see Eqs.~(\ref{epsn}) and (\ref{tnn}), not
between distinct cells. These are motions that correspond to {\em
  optical} phonons.  More formally, we note that the potential energy
terms in Eqs.~(\ref{KG}) and (\ref{KGionic}) are largest in the center
of the Brillouin zone but become completely suppressed at the edge of
the zone. (We remind the reader that Eqs.~(\ref{KG}) and
(\ref{KGionic}) were obtained by expanding around that edge.)
Likewise we note the interaction with acoustic phonons near the edge
of the Brillouin zone, which is represented by the first term in
Eq.~(\ref{omitted}), does not contribute to the random potential at
this level of approximation. Such contributions are however expected
closer to center of the Brillouin zone. For instance, one expects that
thermal expansion accounts in part for the narrowing of the band
gap.~\cite{PhysRev.78.808.2}

We next comment on the requirement of the model from
Refs.~\onlinecite{PhysRevLett.57.1777, PhysRevB.37.6963, 5390028} that
noise-autocorrelation be very strong within a certain distance, $L$,
but decay very rapidly beyond that, as in Eq.~(\ref{CFGaussian}). That
the decay should be so rapid is somewhat puzzling since we expect
spatial correlation in solid to decay no faster than the screened
Coulomb interaction, if that fast. At the same time, the
parameter-free calculation of Grein and John,
Ref.~\onlinecite{PhysRevB.41.7641} and references therein, reproduces
the gross features of the Urbach tail quantitatively at sufficiently
high temperatures, while indicating that non-adiabatic effects are
important, at least for crystalline materials. One may elaborate on
these notions in the present context by first reiterating that the
random-potential model treats the nuclear motions entirely
classically. On the other hand, it is not guaranteed that the local
vibrational configurations corresponding to specific local
fluctuations of the random potential would be in quantum-mechanical
resonance.  Accounting for the presence/lack of such resonances is
routinely done in theories of non-adiabatic charge
transfer.~\cite{OvchEpsOmega} Hereby, one finds the rate of charge
transfer or optical absorption by directly summing over the spectra of
the two sets of lattice-vibrational modes corresponding, respectively,
to the initial and final states of the transferred electron.  As a
result, the charge transfer rate is modulated by wave-function
overlaps between these two sets, i.e., the Frank-Condon factors.

We next outline how to estimate these Franck-Condon factors using the
present picture.  First we note that the spring constant $\kappa$ of a
bond is correlated with the bond strength,~\cite{GrimvallSjodin} hence
one can write qualitatively:
\begin{equation} \label{kappat} \kappa \propto t_\text{e}^\alpha,
\end{equation}
where the transfer matrix $t_e$ is from Eq.~(\ref{tr}) and $\alpha$ is
some positive power, see also Section~\ref{MIT}. Because of the close
association between the bond strength and its stiffness, we expect
that to a localized electron, there will correspond a set of localized
vibrations as well. This can be seen relatively directly as follows.

The vibrational spectrum of a chain corresponding to the electronic
spectrum from Eqs.~(\ref{Huckel})-(\ref{tnn}) is similar to the
electronic spectrum, with the difference that its ground state is at
zero frequency, of course, while the gap between the acoustic and
optical branch spans the vibrational frequency of the weak and strong
bond, respectively, if the masses of the ions are the
same.~\cite{Ashcroft} When the ions masses are different, the
situation is a bit more complicated. Consider the following, 1D
vibrational Hamiltonian:
\begin{equation} \label{Hvibr} \cH_\text{vibr} = \sum_n \frac{m_n}{2}
  x_n^2 + \frac{\kappa_{n, n+1}}{2}(x_{n} - x_{n+1})^2,
\end{equation}
where both the ion mass and the spring constant strictly alternate in
magnitude from site to site:
\begin{equation} m_n = \left\{\begin{array}{l} m_1, n \text{ odd} \\
      m_2, n \text{ even}
    \end{array}
  \right.
\end{equation}
\begin{equation} \label{knn} \kappa_{n, n+1} \equiv \bar{\kappa} +
  (-1)^n \kappa, \hspace{3mm} |\kappa| \le \bar{\kappa}.
\end{equation}
In this case, the vibrational eigenfrequency at wave-vector $k$ is
determined by solving the characteristic equation for the following
matrix:
\begin{equation} \label{TLSvibr} \left( \begin{array}{c|c} - m_1
      \omega^2
      + 2\bar{\kappa} & -2 \bar{\kappa} \cos ka + 2 i \kappa \sin ka \\
      \hline -2 \bar{\kappa} \cos ka - 2 i \kappa \sin ka & - m_2
      \omega^2 + 2\bar{\kappa}
    \end{array}
  \right),
\end{equation}
c.f. Eq.~(22.35) of Ashcroft and Mermin~\cite{Ashcroft} and
Eq.~(\ref{TLS}). Upon introducing the harmonic average of the ion
mass:
\begin{equation} m = \sqrt{m_1 m_2},
\end{equation}
one can re-trace the derivation of Eq.~(\ref{Hcont}) to obtain the
vibrational analog of the continuum Hamiltonian which determines the
eigenvalues for the quantity $m \omega^2$:
\begin{equation} \label{HcontVibr} \cH = - i 2 \bar{\kappa} a \,
  \hat{\sigma}_3 \frac{\prtl}{\prtl x} - 2 \kappa(x) \, \hat{\sigma}_1
  - \frac{\Delta m}{m}\bar{\kappa}(x) \, \hat{\sigma}_2 +
  \frac{M}{m} \bar{\kappa},
\end{equation}
where
\begin{align} \Delta m \equiv m_1 - m_2 \\
M  \equiv m_1 + m_2
\end{align}
and we allow for slow variation in the bond strength. One can make the
Hamiltonian (\ref{HcontVibr}) of dimensions energy by multiplying it
by $a^2$. Analogously to Eq.~(\ref{Hcont}), Hamiltonian
(\ref{HcontVibr}) gives the energy per atomic site.

Although the correspondence between the electronic and phonons
Hamiltonians is not exact---far from it---the similarity is hard to
miss. Note that the mass differential between the two distinct ions
plays a formally similar role for the vibrational spectrum to the role
the electronegativity variation plays in the electronic spectrum. In
contrast with the electronic spectrum, the vibrational spectrum is
coupled not only to long wave-length optical distortions, as embodied
in the $2 \kappa(x) \, \hat{\sigma}_1$ term in Eq.~(\ref{HcontVibr}),
but also to the acoustic modes, via the term $(\Delta
m/m)\bar{\kappa}(x)$.

Localized vibrations are modes whose frequencies fall within the
forbidden gap between frequencies $\omega_v$ and $\omega_c$ in
Fig.~\ref{spectrumSketch}.  The behavior of vibrationally localized
solutions can be understood already from the simplest version of
Hamiltonian (\ref{HcontVibr}), in which the average force constant and
its alternation pattern is spatially homogeneous. The resulting
Klein-Gordon equation for the vibrational eigen-modes is
\begin{equation} \label{KGvibr} \left( - 4\bar{\kappa}^2 \frac{a^2
      \prtl^2}{\prtl x^2} + \Delta_v^2 \right) \psi = \left(m
    \omega^2 - M \bar{\kappa}/m \right)^2 \psi,
\end{equation}
where $\Delta_v$ is the vibrational gap:
\begin{equation} \label{DeltaVibrDef} \Delta_v \equiv [(2 \kappa)^2 +
  ( \bar{\kappa} \, \Delta m/m)^2]^{1/2}.
\end{equation}
The edges of the vibrational gap from Fig.~\ref{spectrumSketch} are
given by the expressions $\omega_{c, v}^2 = M \bar{\kappa}/m^2 \pm
\Delta_v/m$. The maximum achievable phonon frequency is 
\begin{equation} \label{omegaD} \omega^2_D = 2 M \bar{\kappa}/m^2 = 2
  \bar{\kappa}/\mu,
\end{equation}
where $\mu \equiv m_1m_2/M$ is the reduced mass of the bond.

According to Eq.~(\ref{KGvibr}), a localized vibrational mode centered
at $x_0$, would exponentially decay away from $x_0$:
\begin{equation} \psi(x) \propto e^{-|x-x_0|/\xi_v},
\end{equation}
where 
\begin{equation} \label{xiv} \xi_v \equiv a
  \frac{2\bar{\kappa}}{\Delta_v} = a \left(\frac{\kappa}{\bar{\kappa}}
    + \frac{\Delta m}{m} \right)^{-1}.
\end{equation}
This is entirely analogous to how an edge electronic state or midgap
electronic state, in the context of Eqs.~(\ref{Hcont}) and (\ref{KG}),
would decay exponentially rapidly into the
bulk.~\cite{RevModPhys.60.781, PhysRevB.21.2388, ZL_JCP} See also a
pedagogical discussion in Ref.~\onlinecite{LL2}. Likewise, one can
think of the vibrational localization length $\xi_v$ as the
penetration length of an instanton.

To assess the aforementioned Franck-Condon factor, we make the
following mental construct: Imagine an instantaneous aperiodic
configuration of the ions that lead to both a localized electronic
solution and, by virtue of Eq.~(\ref{kappat}), to a localized
vibrational mode. The two localized modes will be centered on the same
site and decay exponentially into the bulk, as just mentioned. The
Franck-Condon factor corresponding to the electron shifted from
location $x_1$ to $x_2$ is proportional to
\begin{equation} \label{FC} e^{-|x_1 - x_2|/\xi_v} \prod_{n}
  e^{-(x_n^{(1)} - x_n^{(2)})^2/2\delta x_v^2},
\end{equation}
where $x_n^{(1)}$ and $x_n^{(2)}$ denote the position of atom $n$
corresponding to the electronic configurations 1 and 2, respectively.
The quantity $\delta x_v$ is the amplitude of the localized
vibrational mode. The vibrational frequencies in question are within
the gap and thus intermediate in value between the acoustic and
optical branch, i.e., roughly half the Debye frequency. Thus $\delta
x_v$ exceeds its zero-point value but not by much. Hence one may
write:
\begin{equation} \delta x_v \sim (2 \hbar a/\xi_v M \omega_D)^{1/2},
\end{equation}
Here we estimated the mass of the mode by counting the ions within the
correlation length $\xi_v$. In view of Eq.~(\ref{kappat}), $\xi_v$
will be comparable to the electronic instanton length, which is
ordinarily of the order 10 lattice spacings in amorphous
semiconductors,~\cite{ZL_JCP, ZLMicro2} but could become quite large
when the forbidden gap becomes small, see Eq.~(\ref{xiv}).

Because of this large mass, the vibrational amplitude $\xi_v$ will be
very small, significantly smaller than the vibrational amplitude of an
individual atom. This implies that the product of individual
Franck-Condon (FC) factors in Eq.~(\ref{FC}) will be very very small
except, perhaps, in the case of strict periodicity of the
lattice. (Note the FC factors for atoms far away from the localized
mode are numerically close to one; the said smallness is largely due
to the atoms contributing to the localized mode.)  The small magnitude
of the FC factor can thought of as the motion of the electron being
virtually recoil-less when it is subjected to a fluctuating potential.
(This lack of recoil on the part of the lattice is reminiscent of the
M\"ossbauer effect.) This, then, suggests that in an internally
consistent treatment, auto-correlation of the random potential should
decay exceptionally fast, by construction. This is consistent with the
postulated rapid decay from Eq.~(\ref{CFGaussian}). Once the
correlation does begin to decay, it should do so like the
Franck-Condon factor in Eq.~(\ref{FC}), i.e., in a Gaussian fashion.
Once again, the large mass of the localized vibrational mode is what
renders the classical treatment of the potential fluctuation from
Refs.~\cite{PhysRevLett.57.1777, PhysRevB.37.6963, 5390028}---while
assuming short-range correlations---internally consistent.

Finally note that the possibility of a somewhat slower decay in the
case of strict periodicity seem to be consistent with the Urbach tail
in crystalline Si exhibiting some structure.~\cite{PhysRevB.41.7641}
No such structure is expected in amorphous materials, as already
mentioned.

\section{The Intrinsic Relationship between Electronic Excitations
  across the Band Gap, Optical Phonons, and Metallicity}

Thus we have argued that in compounds of Type 1 and 2, long wavelength
optical phonons are intrinsically related to and can quantitatively
account for the stabilization of the band gap and, at the same time,
the appearance of the Urbach tail.  The two types of compounds can be
understood as supporting a standing charge density wave whose crests
are centered, respectively, on bonds and ions,
respectively. Significantly, the CDW must have the exact same symmetry
with respect to translations as the lattice itself.  The two types of
CDW are mathematically equivalent, see Eq.~(\ref{Hcont}), but
generally result from very different physical processes. To give rise
to an off-center, bond-order CDW, the atoms would have to move
distances intermediate between the atomic vibrational amplitude, $\sim
a/10$ and atom spacing, $a$. An example of such a process is the
simple-cubic to rhombohedral arsenic transition.~\cite{silas:174101,
  PhysRevB.77.024109} For an on-site CDW, on the other hand, two
following possibilities come to mind: On the one hand, such a CDW can
be created by placing two atoms with differing electronegativies in
the alternating positions on a bi-partite lattice such as in the NaCl,
CsCl, wurzite, or zincblende lattice. In addition, a spontaneous
polarization can be envisioned, such as during charge
disproportionation,~\cite{ATTFIELD2006861} which may or may not be
accompanied by a displacive transition.

\begin{figure}[t]
\center
\includegraphics[width =  \figurewidth]{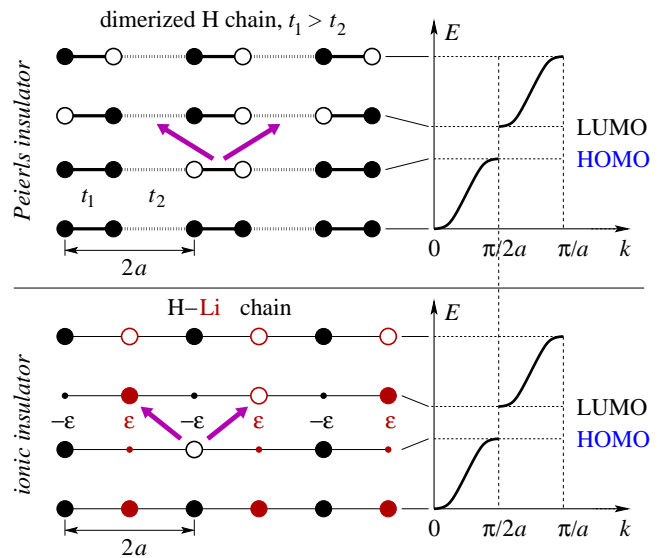}
\caption{\label{MO} A molecular orbital (MO) representation of the
  wavefunctions corresponding to the band edges in a 1D insulating
  chain. The top and bottom panel correspond with, respectively, Type
  1 and 2 insulators. (In the Type 1 case, we adopted the extreme case
  $\varepsilon=0$.) Filled and empty circles correspond with opposite
  sign of the wavefunction, on the respective site. The size of the
  circle reflects the local magnitude of the wavefunction. The arrows
  show how the charge is transferred during a HOMO $\to$ LUMO
  excitation.}
\end{figure}

The apparent connection between the presence of the optical gap, on
the one hand, and optical phonons, on the other hand, can be discussed
using already simple notions of the molecular orbital (MO) theory.
The band edge MOs for compounds of type 1 and 2 are sketched in
Fig.~\ref{MO}, top and bottom panels, respectively. There, one can
trace local polarization---spontaneous or that arising upon optical
excitation---to movements of either the lobes or nodes in the
corresponding electronic wavefunction. Clearly, such local
polarization patterns correspond to an optical phonon.

Compounds of Type 3 may seem to not fit in the present formalism at a
first glance. Consider, however, crystalline silicon or germanium,
which adopt the diamond lattice at normal conditions. This lattice has
two atoms in its primitive cell. The respective sets of these two
equivalent atoms can be thought of as each forming equivalent FCC
sub-lattices. (The two FCC sub-lattices are shifted relative to each
other along the main diagonal of the cubic unit cell, by a quarter of
the length of the diagonal.) The primitive cell of the diamond lattice
is the same as that of either of the FCC sub-lattices. The overall
electronic structure can be thought of, even if somewhat
unconventionally, as a resonance between two sets of atomic orbitals
centered, respectively, on the two FCC sub-lattices.  Yet each of
these sets corresponds to a charge density wave that has the same
primitive cell as the lattice itself.
Thus the valence and conduction band of Si can be thought of,
respectively, as the bonding and anti-bonding combination of two CDW
states. Similarly to the simple picture in Fig.~\ref{MO}, excitations
across the gap or spontaneous polarization will involve charge
redistribution within the primitive cell of the lattice---and hence
coupling to optical phonons---because an anti-bonding orbital has an
extra node compared with the corresponding bonding one. The so
locally-polarized state would have the same symmetry with respect to
translations as the lattice itself, just as it was for materials of
Type 1 and 2. (We anticipate the detailed pattern of local
polarization could be a quite more complicated than that sketched in
Fig.~\ref{MO} because the constituent CDW states are themselves
complicated. Consistent with this notion, silicon and electronically
similar materials have an indirect gap.)

Conversely, suppose there exists an insulator that has just one atom
per primitive cell. The preceding discussion suggests that excitations
across the gap could {\em not} create a local-polarization pattern
that has the same translational symmetry as the lattice. This would be
fine for a solid made of closed shell atoms: For instance, optical
transitions can be simply thought of as optical excitations of
weakly-interacting individual atoms. Consistent with this, elemental
helium metalizes only at high
pressures.~\cite{PhysRevLett.112.055504} However in the presence of
covalent bonding---i.e. bonding between partially filled atoms
shells---this picture seems potentially problematic because every
valence atomic orbital is strongly hybridized with the orbitals
centered on neighboring atoms, by the very definition of covalent
bonding. It stands to reason that optical excitations {\em would}
create a polarization pattern in a covalently bonded solid.  Yet the
pattern would have to have a lower translational symmetry than the
lattice itself. Thus polarization-induced stabilization and broadening
of the band edge, if any, would be much weaker---despite the molecular
orbitals being delocalized!---suggesting the supposition that our
insulator has just one atom per primitive cell was
internally-inconsistent.

To make the above notion more systematic, we first recall a formal,
yet fundamental aspect in which insulators are different from metals:
First imagine a set of molecular orbitals (MOs) that diagonalize an
effective one-electron Hamiltonian. The totality of the MOs form a
complete basis set and so the atomic orbitals can be recovered from
the MOs by a suitable unitary transformation. Now suppose the MOs are
not all filled, as would be the case if the participating atoms are
not all closed-shell. The filled MOs alone no longer comprise a
complete set. Thus, the Mullican-Hund (MO) and Heitler-London
(localized) descriptions are not equivalent unless all MOs are filled,
as emphasized by Burdett.~\cite{Burdett1995, BurdettMetInsTrans} This
elementary notion naturally leads to a more general observation that
the band description and the description in terms of localized Wannier
orbitals are equivalent only in insulators. In fact, it was shown
relatively recently that there do exist exponentially localized
Wannier orbitals for 2D and 3D
insulators.~\cite{PhysRevLett.98.046402} On an historical note,
localized-MO formalisms have attracted attention of physicists and
chemists alike over the years,~\cite{RevModPhys.32.296,
  RevModPhys.35.457, PhysRev.181.25, PhysRevLett.21.13,
  doi:10.1063/1.1677859, GHL} despite their non-uniqueness. Perhaps
the most immediate use of localized Wannier-like orbitals is to
represent complicated results of quantum-chemical calculations so as
to appeal to the concepts of the canonical two-center bond, closed
shell interactions, and lone pairs. Conversely, this formalism helps
highlight situations where bonding becomes truly
multi-center.~\cite{GHL}

Now imagine a covalently bonded insulating solid made of partially
filled atomic orbitals. For the sake of completeness, assume there is
one filled and one empty band. The orbitals comprising the filled band
can be represented in terms of mutually-orthogonal Wannier orbitals,
call them the ``electron Wannier orbitals.'' Likewise, the orbitals
comprising the empty band can be represented in terms of
mutually-orthogonal ``hole Wannier orbitals.'' The charge distribution
pertaining to the electron Wannier orbitals(but not the individual
orbitals!)  must have the symmetry of the lattice with respect to
translations, and likewise for the hole Wannier orbitals. At the same
time, the electron and hole Wannier orbitals are not identical;
instead, they are mutually complimentary.  The former set is made of
the bonding orbitals, and the latter set of the anti-bonding
orbitals. Thus we conclude that in a covalently bonded insulator,
excitation across the forbidden gap must result in a polarization
pattern that has the same symmetry with respect to translations as the
lattice itself. At the same time, the charge redistribution that leads
to this polarization pattern must involve charge transfer between
atomic sites and/or interatomic spaces, analogously to the specific
examples considered in the beginning of this Section. Indeed, the
Wannier orbitals are not the same as the atomic orbitals. Already the
number of the former is less than that of the latter.  Thus a Wannier
orbital covers more than one atomic orbital. Furthermore, at least a
subset of the Wannier orbitals in a covalently bonded system must
include combinations of orbitals centered on {\em distinct} atoms, as
opposed to orbitals centered on the same atom. To drive this notion
home we recall that the Wannier orbitals are linear combinations of
the occupied (vacant) MOs, not the atomic orbitals themselves. The
latter, informally speaking, are not even accessible by themselves but
only in bonding (anti-bonding) combinations of atomic orbitals
centered on distinct atoms. Many specific examples of such localized
orbitals can be found in the literature.~\cite{RevModPhys.32.296,
  RevModPhys.35.457, doi:10.1063/1.1677859, PerkinsStewart1980,
  Levine09, GHL}

Thus we conclude that in a covalently bonded crystalline insulator,
the polarization pattern resulting from an excitation across the band
gap can have the symmetry of the lattice itself and, at the same time,
involves charge transfer between {\em distinct} atoms and/or
interatomic spaces. Thus the primitive cell of a covalently bonded
insulator must contain at least two distinct atoms and thus should
exhibit optical phonons. In the case of ionic or aperiodic solids,
this is automatically true, of course. (We note that liquid insulators
would be classified as amorphous solids because electrons move much
faster than nuclei.) A corollary of this observation is that if a
solid lacks optical phonons, it must be a metal. Ordinarily, a solid
lacking optical phonons has no more than one atom per primitive cell
(but see below for additional discussion).  These notions are
consistent with experience: For instance, browsing through the
Periodic Table reveals that every elemental solid that is not metallic
has at least two atoms per primitive unit. In some cases, we have
covalently bonded solids such as silicon or germanium; the diamond
lattice has two atoms per primitive cell. In the case of halogens or
nitrogen, for instance, the solids can be thought of as molecular
solids made of closed-shell dimers. Either way, that these molecular
solids are insulating is consistent with the present discussion. To
avoid confusion we note that the reverse of this conclusion is
generally not true: The presence of optical phonons does not preclude
the solid from being metallic.

Although the above argument was stated using atomic orbitals as a
local basis set, it can be reformulated without explicit reference to
atomic orbitals but, instead, only using distinct sites in space. This
would be useful to systems such as Wigner solids, in which the
counter-charge is distributed uniformly in space by construction and
so there is no obvious set of atomic-like basis set functions, see
below. Stated more generally, the preceding conclusion can be stated
as follows: The primitive cell of an insulating solid should contain
at least two spatially inequivalent localized basis orbitals. This
could be realized in a variety of ways. When atoms are indeed present,
one can imagine a solid with just one atom per primitive cell but a
CDW that has a lower symmetry. Imagine for instance a bcc lattice made
of identical ion cores that hosts an on-site CDW whose symmetry is
lower, namely, that of the CsCl lattice. The latter lattice is
bi-partite, of course, consisting of two equivalent simple cubic
lattices.  Likewise, an elemental solid exhibiting the simple cubic
structure would have to be metallic but could become insulating upon
emergence of a CDW whose crests and troughs each form the rock salt
structure; the latter represent a bi-partite lattice made of two fcc
sub-lattices. One can also imagine various distortions of the above
lattices, such as tetragonal or monoclinic, that do not lead to an
increase in the number of atoms per primitive unit.

Such symmetry-lowered bi-partite solids would host at least three
optical phonon branches even though the original ions comprise a
higher symmetry lattice. X-ray or electron scattering would pick up
this lowering of the symmetry (unless the CDW is dynamic on the time
scale of the experiment). In this sense, these solids should not be
regarded as containing only one atom per primitive cell despite the
ionic cores comprising the sub-lattices of lower symmetry being
identical. Instead, one has a symmetry breaking in the way of charge
disproportionation that effectively leads to the appearance of two
chemically distinct species. We are not aware of charge
disproportionation in the form of simple bi-partite lattices in {\em
  elemental} compounds, however more complicated charge
disproportionation patterns have indeed been observed in the form of
host-guest structures.~\cite{PhysRevB.77.024109} The emergence of such
patterns does not seem to affect the fact of metallicity of the
material.  Perhaps bi-partite like charge disproportionation patterns
could be observed at sufficiently high pressures. In this regard,
studies on the high pressure behavior of hydrogen come to
mind.~\cite{doi:10.1063/1.3679751} These studies suggest that solid
hydrogen would become metallic {\em before} its structure would be so
symmetric as to contain only one atom per primitive cell, see also
Ref.~\onlinecite{PhysRevLett.120.255701}. No charge dispropotionation
seems to occur.

We have mentioned the zincblende and wurzite lattices as examples of
standing charge density waves. Here we note that their ``parent
lattices,'' i.e. the the diamond and lonsdaleite lattices,
respectively, for instance, already have two atoms per primitive cell
and do not have to be metallic.

Even in the absence of charge disproportionation, the symmetry can
still be lowered by the emergence of spin-density wave (SDW). This
symmetry lowering would not be picked up by X-ray diffractometry, but
could be detected by neutron scattering (unless the SDW is dynamic on
the time scale of the experiment.) We do not expect that a
spin-density wave by itself would cause the material to exhibit
optical phonons let alone become insulating. For instance, chromium is
known to spontaneously develop an SDW at low temperatures---both
commensurate and incommensurate.~\cite{RevModPhys.60.209} As a
secondary effect stemming from electron-phonon interaction, a strain
wave~\cite{ISI:A1978FY89000009} and a CDW~\cite{PhysRevB.13.295}
emerge alongside as well. Neither of these lead to the disappearance
of the metallicity.  In discussing purely electronic mechanisms to
lower the symmetry of the lattice, we should mentioned that, in the
first place, the emergence of charge/spin density wave or any kind of
charge disproportionation could lead a more basic kind of symmetry
breaking, namely, in a way of static lattice
distortions.~\cite{ATTFIELD2006861} Perhaps the simplest example of
such a symmetry broken solid is the rhombohedral
arsenic,~\cite{silas:174101} which can be thought of as 3D analog of a
Peierls insulator.~\cite{BurdettLeePeierls} In the case of such
displacive transitions, optical phonons will emerge automatically.  An
insightful way to connect various lattice distortions with the
underlying electronic phenomena is through notions of reduction of
dimensionality,~\cite{PapoianHoffmann2000} again consistent with the
notion of the metal-insulator transition as a symmetry lowering.

Yet in some cases a spin-density wave could be in fact indicative of
substantial---even complete---electronic localization. According to
existing studies, a periodic Wigner solid in 3D should be
bcc.~\cite{Fuchs585, doi:10.1063/1.1703670} The Wigner bcc solid has
just one site per primitive unit implying it must be metallic even at
absolute zero. This would seem to contradict the premise of the
electron assembly being a solid, in which the electrons are strictly
localized. The latter point deserves some elaboration.  First we note
that the reported phase diagrams are most likely oversimplified in
that they show a sharp transition line between the liquid and solid
phase in the temperature-density plane. Yet the transition must be
discontinuous except, perhaps in one isolated point.~\cite{L_AP,
  LandauPT1, LandauPT2, Brazovskii1975} Thus, properly, those phase
diagrams should show a substantial {\em coexistence} region, which
physically corresponds to macroscopic parts of the system being
occupied by the solid and liquid phase, respectively. This means that
the reported transition density is likely higher than the density of
crystal and lower than the density of the liquid that would actually
spatially coexist. (The reported transition itself probably
corresponds with mechanical, not thermodynamic
melting.~\cite{L_Lindemann})
Now, imagine approaching the solid-to-liquid transition from the
crystal side, all at $T=0$ (!), and suppose for a moment that instead
of spatial coexistence, the appearing liquid is uniformly distributed
over the sample. (This would imply the solid is metallic.) Clearly
such uniform liquid would be of much lower density than required for
its stability. Hence, the crystal and the liquid cannot coexist in the
same region but only distinct regions in space.  This notion adds to
the observation due to Golden et al.~\cite{GHL} who have suggested
electronic fluids with differing degrees of localization would not mix
at sufficiently high density. To avoid confusion we note that at
finite temperatures, some of the electrons are allowed to be promoted
to the conduction band subject to activation across the forbidden
gap. The density of this (thermally activated) electronic fluid could
be made arbitrarily small since it is not in bulk equilibrium with the
solid.

At the same time, the possibility of the Wigner crystal hosting a CDW
with the translational symmetry of the CsCl lattice can be excluded
because of quantum fluctuations: The system would readily minimize its
kinetic energy by assuming a superposition of charge-disproportionated
configurations.

The only remaining way to lower symmetry so as prevent the primitive
unit to have more than one inequivalent site is to have an
antiferromagnetic (AF) ordering. This is consistent with the results
of quantum Monte Carlo simulations.~\cite{PhysRevB.70.094413} In 3D,
the simplest kind of AF ordering could be realized by having the Cs
sites in the CsCl lattice to be polarized up and Cl sites polarized
down, for instance. In 2D, such anti-ferromagnetic ordering could be
realized by an alternating sequence of chains polarized up and
down.~\cite{PhysRevLett.102.126402}

In a distinction from insulating solids made of atoms, the Wigner
solid exhibits only three phonons branches, two of which are gapless
and correspond with the transverse acoustic phonon branches of a
regular solid; neither incur local density changes.  On the other hand
the longitudinal mode, which represents the plasma oscillations, is
gapped. The gap comes about because longitudinal sound waves do incur
density changes while the Coulomb forces are long range thus imposing
a finite energy cost even on the longest wavelength
vibrations.~\cite{PinesEXS} The plasmon can be thought of as a Higgs
boson.~\cite{AndersonBasicNotions, PhysRev.130.439}

\section{Metal-Insulator Transition as a Symmetry Breaking}
\label{MIT}

We have discussed manifestations of an intrinsic connection between
the presence of a band gap and optical phonons.  Here we wish to take
a more general perspective on this connection. The dimerized chain,
such as that depicted in Fig.~\ref{MO}, is clearly of lower symmetry,
with respect to translation, than the original non-dimerized
chain. But at the electronic level, the symmetry breaking is actually
a bit more interesting. Indeed, consider the $k=\pi/2a$ orbitals
centered on the odd- and even-numbered sites from Eq.~(\ref{odd}),
call them $\psi_1$ and $\psi_2$. These have the same energy if
$t=\varepsilon=0$. Any pair $\psi_1 \cos(\phi/2) + i \psi_2
\sin(\phi/2)$ and $i \psi_1 \sin(\phi/2) + \psi_2 \cos(\phi/2)$ of
these two orbitals would form an equivalent set of solutions as well,
for {\em any} value of $\phi$. Yet this (continuous) symmetry with
respect to $\phi$ is broken following the dimerization. For instance,
for the ``Peierls insulator'' in Fig.~\ref{MO}, $\varepsilon=0$ the
value $\phi=\pi/2$ is selected, as is easy to see from
Eq.~(\ref{TLS}). The ionic case, on the other hand, selects out
$\phi=0$.

There are other ways to look at this symmetry breaking: The optical
phonons can be thought of resulting from chemical ordering, in the
form of a specific arrangements of ions or bond strength. To restore
the symmetry in the latter case, the ions or bonds of differing
strength would have to be able to swap places, which would require
some nuclear motions, too. The strength of the CDW appears to be an
appropriate order parameter to describe the symmetry breaking.  Viewed
this way, the optical phonons could be thought of as fluctuations of
the order parameter of the metal-insulator transition.  These notions
prompt us to take a field-theoretic perspective on the discussion from
the preceding Sections. As an additional benefit, a symmetry based
discussion is expected to be relatively robust with regard to
many-body effects.


We will build a density functional description in the spirit of
Landau's analysis of the liquid-to-solid transition,~\cite{LandauPT1}
where the order parameter reflects the magnitude of a density wave
near the wavelength corresponding to the particle spacing in the
putative solid, see also Ref.~\onlinecite{L_AP}. It is most
convenient, for the present purposes, to use the parameters
$\varepsilon$ and $t$ from Eq.~(\ref{Hcont}) as the order
parameters. These can be regarded as the components of the molecular
field~\cite{L_AP} that control entries of the electronic density
matrix
\begin{equation} \label{DM} P_{\alpha \beta} = \sum_{s = \pm 1/2} \la
  c^\dagger_{\alpha, s} c_{\beta, s}^{} \ra = 2 \sum_{k, \,
    \text{occ}} \psi_{k \alpha}^{} \psi_{k\beta}^{*},
\end{equation}
Specifically, the spatial variation of the on-site electronic density:
\begin{equation} \label{rho} \rho \equiv \frac{1}{2}(P_{2n+1, 2n+1} -
  P_{2n, 2n}).
\end{equation}
can be computed as the following derivative:
\begin{equation} \label{rhoeps} \rho = - \frac{\prtl E}{\prtl
    \varepsilon}.
\end{equation}
Likewise, the variation in the bond strength
\begin{equation} \label{sigma} \eta \equiv \frac{1}{2}(P_{2n,
    2n+1} - P_{2n, 2n-1}) + \text{C.C.}
\end{equation}
is the conjugate variable to the transfer integral:
\begin{equation} \label{sigmat} \eta = - \frac{\prtl E}{\prtl t}.
\end{equation}
By adopting the restrictions from Eqs.~(\ref{epsn}) and (\ref{tnn}) we
limit ourselves, by construction, to wavelengths near the period of
the dimerized chain. To obtain a density functional theory in the
sense of Kohn and Sham,~\cite{PhysRev.140.A1133}, which would operate
on the variables $\rho$ and $\eta$, one can perform a Legendre
transform,~\cite{GHL} something we will not do here.

Already a quick look at the spectrum in Eq.~(\ref{spectrumEq}) shows
that the energy $E$ of the non-interacting model, at half-filling, is
maximized at $\varepsilon = t = 0$ and monotonically decreases away
from the maximum. Near $\Delta=0$, is expected to behave smoothly in
spatial dimensions higher than one:
\begin{equation} \label{E2} E = \frac{a_2}{2} \Delta^2 + \text{higher
    order terms},
\end{equation}
where the coefficient at the quadratic term 
\begin{equation} a_2 \sim \frac{1}{W}
\end{equation}
is largely determined by the width $W$ of the valence zone. In 1D, the
coefficient $a_2$ is actually negative and singular $a_2 \propto \ln
\Delta^2$,~\cite{PhysRevB.22.2099} implying that irrespective of
interactions, the chain will dimerize. Indeed, fixing $\varepsilon$
and adding a penalty for distorting the bonds, $\propto (+ t^2)$ will
produce a bistable potential energy surface, if $\varepsilon$ is not
too large. The two minima correspond to the alternative ways to
dimerize. In dimensions higher than one, the non-interacting part is
no longer singular; the symmetry breaking may or may not take place
depending on the geometry and interactions. For instance,
electron-electron repulsion will generally suppress dimerization.

Above said, let us assume until further notice that the system does
indeed exhibit a {\em continuous} symmetry breaking:
\begin{equation} a_2 < 0,
\end{equation}
just like the non-interacting 1D models from the preceding Sections.
Because the order parameter is vectorial, the broken symmetry is
continuous, consistent with the elementary discussion in the beginning
of this Section. Cases of broken continuous symmetry are interesting
because they sometimes result in massless, Goldstone modes. Let us see
that no such modes arise in the present case however.

\begin{figure}[t]
  \centering
  \includegraphics[width = 0.45 \figurewidth]{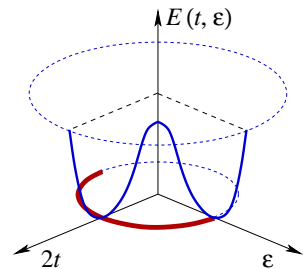}
  \caption{\label{mexhat} A hypothetical, rotationally-symmetric free
    energy profile as a function of the strength of the charge-density
    wave, $\rho$ and $\eta$ standing for the on-site and off-site
    components, respectively. The semi-circle at the bottom of the
    potential exemplifies a hypothetical trajectory in the $(\rho,
    \eta)$ space for a transition between two distinct CDW states.}
\end{figure}

\begin{figure}[t]
  \centering
  \includegraphics[width = 0.7 \figurewidth]{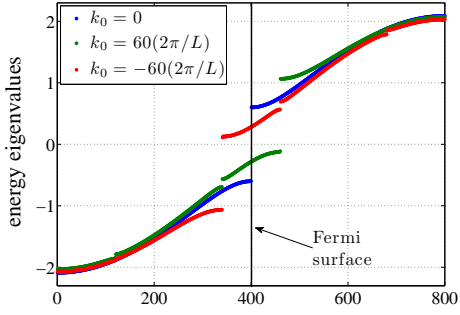}
  \caption{\label{twist} The spectrum of the model (\ref{Huckel}) with
    a periodically varying $\varepsilon = \Delta_0 \cos(k_0 x)$ and $t
    = (\Delta_0/2) \sin(k_0 x)$. $\bar{t} = 1$, $\Delta = 0.6$, N =
    800.}
\end{figure}

First suppose the opposite, i.e., that the broken symmetry functional
has the ``mexican hat shape'' sketched in Fig.~\ref{mexhat}. The
Goldstone mode, if any, would correspond with fluctuations of the
vectorial order parameter such that its end moves along the local
direction of the circular groove. In contrast with such low magnitude
motions, one can also envision motions where the end of the vector
moves a substantial amount. Such motions would correspond to a
crossing of a domain wall separating equivalent symmetry-broken
states, such as states with distinct polarizations of the Heisenberg
magnet. In the present context, these distinct states would correspond
to equivalent CDW states. Yet in the absence of magnetic field,
self-generated or externally imposed, such a trajectory would be
forbidden. Indeed, consider the Hamiltonian (\ref{Hcont}) where the
$x$-axis is perpendicular to the domain wall and the vector $[2t(x),
\varepsilon(x)]$ performs a circular motion at a steady rate
alongside: $\Delta(x) =\Delta_0$, $\prtl \varphi/\prtl x = k_0$. The
sign of the constant $\varphi'_0$ depends on the direction of
rotation. This yields the following spectrum:
\begin{equation} E_k =  -\bar{t} a k_0 \pm \sqrt{\Delta_0^2 +
    (2 \bar{t} a k)^2}.
\end{equation}
That is, the Fermi level shifts up or down depending on the sign of
$k_0$. This shift contradicts the notion of the two directions of
${\bm k}_0$ being physically equivalent. Less obvious from the
continuum solution is that a non-vanishing $k_0$ makes the system {\em
  metallic}, thus contradicting the original premise of the CDW being
a gapped insulator. Indeed, for each additional ``revolution'' of the
$(\varepsilon, 2t)$ vector, a state is transferred between the valence
and conduction band, the direction of transfer depending on the sign
of $k_0$, see the solution of a discrete system in
Fig.~\ref{twist}. There we also observe that additional, relatively
narrow forbidden gaps may appear in the spectrum. 

We note it is in principle possible to have a sense of direction for
an interface in the CDW phase, if the time-inversion symmetry is
broken. The latter would be the case, for instance, when (local)
magnetic field is present. Such local magnetic fields could arise, for
instance, in a suitably modified Hubbard model from vortex-like
structures made of spins; the structures extend in a 2D
plane.~\cite{PhysRevB.69.224515} When standalone, these so called {\em
  meron} vertices are infinitely-costly and thus end up forming
strongly bound meron-antimeron pairs, similarly to extended
dislocations.

\begin{figure}[t]
\center
\includegraphics[width = .7 \figurewidth]{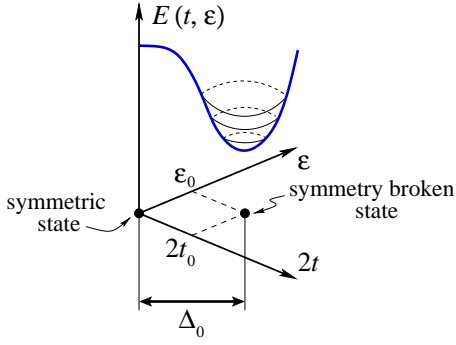}
\caption{\label{sketch} Typical behavior of the optical absorption
  spectrum as a function of temperature. The width of the exponential
  tail of localized states is seen to change synchronously with the
  mobility edge.}
\end{figure}

We thus conclude that the free density functional can not have minima
with vanishing curvature but, instead, should exhibit isolated minima
separated by barriers. Such a minimum can be described, near its
bottom, by a positively-defined quadratic form:
\begin{align} \label{Eteps} E = & \frac{\alpha_{1}}{2} (t-t_0)^2 +
  \frac{\alpha_{2}}{2} (\varepsilon-\varepsilon_0)^2,
\end{align}
where for simplicity we assumed that the cross-term vanishes. (This
can be always achieved by an appropriate rotation.) The (bulk) energy
function is sketched in Fig.~\ref{sketch}. Such a stable minimum can
be obtained by using a higher order term with a positive coefficient
in Eq.~(\ref{E2}); a cubic or quartic term will ordinarily
suffice. Simple algebra shows that the coefficients $\alpha$ are equal
to $|a_2|$ times a numerical factor of order one:
\begin{equation} \label{alphaW} \alpha_i \sim 1/W.
\end{equation}

A proper field theory is obtained by adding to Eq.~(\ref{Eteps}) a
penalty for spatial variation of the parameters $t$ and $\varepsilon$,
for instance in the form of square gradient terms $K_t (\nabla t)^2/2$
etc. The ``spring constant'' $K_t$ and its likes would be determined
by the detailed interactions and geometry; they are complicated
objects that that we have not discussed in this article. While the
value of $t$ can be thought of as controlled by the propensity to
displacive transitions, penalty for its spatial variation seems to
require the knowledge of higher order correlations.  At any rate, the
energy function (\ref{Eteps}) corresponds to a gapped spectrum since
$k\to 0$ fluctuations in the quantities $t$ and $\varepsilon$ incur a
finite cost. This cost, and hence the gap, are proportional to the
quantities $\alpha_i$. Combined with Eqs.~(\ref{Hcont}) and
(\ref{HcontVibr}), and upon inclusion of square-gradient terms,
Eq.~(\ref{E2}) constitutes a field theory. We stress that the negative
sign of the coefficient $a_2$, if any, would come about automatically
in a self-consistent treatment. Once $a_2 < 0$, we arrive at a
symmetry broken version of Eq.~(\ref{E2}), such as that given in
Eq.~(\ref{Eteps}).

To perform a basic check on the internal consistency of this
description we recall that by construction, out-of-phase motions of
the ions comprising the primitive cells will cause variation in the
parameters $t$ and $\varepsilon$. Denoting the amplitude of this
motion with $\delta r$, one obtains, per particle:
\begin{equation} E = \frac{1}{2}[\alpha_1 (\prtl t/\prtl r)^2 +
  \alpha_2 (\prtl \varepsilon/\prtl r)^2] (\delta r)^2,
\end{equation}
The quantities $(\prtl t/\prtl r)^2$ and $(\prtl \varepsilon/\prtl
r)^2$ correspond with $(V/a)^2$ from Eqs.~(\ref{Vr}) and
(\ref{Vlogr}). This notion, together with Eq.~(\ref{alphaW}), yields a
simple qualitative expression:
\begin{equation} E = \frac{V^2}{2Wa^2} (\delta r)^2.
\end{equation}
At the same time, a spatially homogeneous $\delta r$ corresponds with
the $k=0$ optical phonon. Equating the above expression with the
kinetic energy of the vibrational motion of a bond at the Debye
frequency from Eq.~(\ref{omegaD}), one obtains
\begin{equation} \mu \omega_D^2 \sim \frac{V^2}{Wa^2}.
\end{equation}
This can be profitably rewritten by noting that $\omega_D a$ gives the
speed of sound $c_s$, up to a numerical factor of order one, while
$(\mu/a^3) c_s^2$ is the elastic constant,~\cite{LLelast} call it $K$,
again up to a factor of order one. Thus,
\begin{equation} K a^3 \sim \frac{V^2}{W}.
\end{equation}
The l.h.s. usually is of the order of a few electron volts, in actual
materials.~\cite{LW, LW_RMP} The r.h.s. is also of the order of a few
eVs, an encouraging sign. Note also that the scaling above is
consistent with Eq.~(\ref{kappat}). The above equation suggests the
elastic properties of a solid are related in a simple way to the
(logarithmic) derivative of the on-site energies and transfer
integrals with respect to the bond length, on the one hand, and the
band width, on the other hand.

Besides highlighting this relation, the present field theory drives
home the intrinsic connection between the fluctuations of the order
parameter for the metal-insulator transition and the long-wavelength
optical phonons. Thus the temperature-driven increase in the band gap
can be thought of as caused by the increasingly large fluctuations in
this order parameter.

Further insight can be obtained by adopting a specific functional form
for the functional. For simplicity, we explicitly treat only the
longitudinal component of the order parameter. (The transverse mode is
gapped, too!) Assuming the higher order terms is a quartic:
\begin{equation} E = \frac{a_2}{2} \Delta^2 + \frac{a_4}{4} \Delta^4,
  \hspace{3mm} (\Delta \ge 0),
\end{equation}
one obtains that $\Delta_0^2 = a_2/a_4$ and, consequently, that
\begin{equation} \omega_D^2 \propto \frac{1}{W} \propto \Delta_0^2 .
\end{equation}
This relation explicitly demonstrates that the finite excitation gap
for the optical phonon is directly tied to the non-zero expectation
value of the order parameter. One notices at least a superficial
similarity with the field theory of electroweak interactions, where
the mass of the (vector) W and Z bosons is determined by the
expectation value of the Higgs field. In the present context, the role
of the vector bosons is played by the optical phonons while the
forbidden gap $\Delta$ is the Higgs field itself. We further note that
the longitudinal polarization of each of the vector boson ``eats up''
the Goldstone mode that would have been generated as a result of
symmetry breaking; this we saw explicitly when discussing the putative
picture in Fig.~\ref{mexhat}. This notion seems consistent with the
general observation that when broken symmetry is not only continuous,
but is also a gauge symmetry, Goldstone modes ends up being ``eaten''
by longitudinal components of massive fields.~\cite{Peskin} We note
that the dimerization does represent a lowering of a gauge symmetry
since acoustic phonons are gauge particles.~\cite{KLEINERT198351}

No such similarity is obvious for discontinuous transitions, however.
In this case, the simplest form of the functional would be~\cite{L_AP,
  LandauPT1}
\begin{equation} E = \frac{a_2}{2} \Delta^2 + \frac{a_3}{3} \Delta^3 +
  \frac{a_4}{4} \Delta^4, \hspace{3mm} (\Delta \ge 0)
\end{equation}
where now $a_2 > 0$ and $a_3 <0$. Such discontinuous transitions would
be the norm rather than exception in systems with harsh close-range
repulsion since the third order terms is favored by closed-packed
motifs.~\cite{L_AP, PhysRevLett.41.702, AndersonBasicNotions}
Elementary calculations show that neither the location $\Delta_0$ nor
the curvature of the bulk term at $\Delta_0$ are simply related to the
band width $W$. The lack of an explicit connection in this case is,
perhaps, not too surprising. For instance, the insulator-to-metal
transition in Si or Ge coincides with the {\em melting} of these
solids whereby the coordination number increases from 4 to roughly
6.~\cite{Mott1993} The liquid-to-solid transition is particularly
challenging to describe systematically because the elementary
excitation above and below the transition are so
different:~\cite{Lfutile, AndersonBasicNotions} In the solid, the
appropriate degree of freedom is the bond vibration (the phonon) while
in the liquid, it is density fluctuations.

\section{Concluding Remarks}

We have established an intrinsic connection between the presence of an
insulating gap and the existence of optical phonons. The reason is
that in the presence of non-closed shell bonding, electronic
excitations across the gap and local polarization both involve charge
transfer among atoms and inter-atomic spaces. This necessitates that
the primitive cell of a crystalline solid should have at least two
atoms, hence, have optical phonons. By some accounts, the distinction
between acoustic and optical phonons is only formal and has to do with
how one book-keeps the vibrational modes. Yet we have seen the
presence of a gap in the vibrational spectrum is significant in that it
signals a lowering of the symmetry.

While in the presence of optical phonons, the material could be either
a metal or insulator, their {\em absence} necessarily implies the
material must be a metal. A notable exception is weakly bound
condensates made of closed shell atoms. In this case, excitations
across the gap do not have to involve charge transfer between distinct
atoms. Another arguable exception is the Wigner crystal, which
nonetheless must have a lower symmetry than that of the bcc lattice,
at least transiently, by exhibiting antiferromagnetism.

The existence of the optical phonons leads to a wealth of phenomena,
such as the temperature dependence of the optical gap and the width of
the Urbach tail of localized states below the mobility edge, in
crystalline and amorphous insulators alike. These phenomena are
intrinsically related and were the original motivation for the present
study.

The essential role of symmetry-breaking in the emergence of the
optical phonons prompted us to take a field-theoretical perspective on
the temperature dependence of the insulating gap and its appearance in
the first place. We have sketched the structure of a field theory, in
which the role of the order parameter is played by the insulating gap
itself. The phonon frequency scales with the expectation value of the
gap, implying a common cause. (In metals exhibiting optical phonons,
interactions underlying this common cause are also present, of course,
but do not lead to the metal-insulator transition.) The
temperature-driven narrowing of the gap simply corresponds with the
temperature driven increase in the fluctuations of the order
parameter.

For the present field theory, we wrote down what are essentially
non-interacting Hamiltonians for the electrons and phonons. The
various interactions are encapsulated in the part of the functional
that governs the spatial distribution of the band gap. This part is
highly non-linear and generally very complicated. In the present
treatment, much of this non-linearity comes about because of the
nuclear motions and the possibility of displacive instabilities.  In
actuality, electronic interactions should be included more explicitly
since these generally affect the electronic motions even when the
nuclei and insulating gap are fixed.

Already at the level of the simple molecular orbital theory, the
metal-insulator transition is seen as a breaking of continuous
symmetry. Yet one can show explicitly that such breaking does not lead
to the emergence of Goldstone modes. It is interesting that during the
transition from delocalized to localized electronic states for random
Hamiltonians of the type in Eq.~(\ref{Hrandom}) the broken symmetry is
also continuous and that the Goldstone theorem is violated,
too.~\cite{MCKANE198136} In the latter description, the order
parameter is the electrical conductivity. We have seen that in common
materials, both a localized tail of states and a (stabilized) mobility
both require the presence of optical phonons in the first place. The
emergence of the optical phonons thus seems to supersede the physics
of Anderson localization in the metal-to-insulator transition, except,
perhaps, in specially engineered ``dirty'' systems. Perhaps, most
common examples of such dirty systems are {\em photonic} insulators,
such as milk. In any event, such random Hamiltonians are still an
invaluable way to approach the question of the functional form of the
tail of localized states.~\cite{PhysRevLett.57.1777, PhysRevB.37.6963,
  5390028} We have speculated that the concomitant localization of the
vibrational modes causes a very abrupt drop-off in the
auto-correlation for the effective random potential acting on the
electrons. Let us add here that it seems likely that this vibrational
localization also explains why the configurational entropy of strongly
interacting liquids can be often quantified by calibrating the entropy
of the liquid by the entropy of weakly interacting liquids made of
closed-shell molecules.~\cite{LW_soft}

We hope that the present take on the metal-insulator transition, in
conjunction with structural degrees of freedom, will be useful in the
area of {\em aperiodic} charge density waves. A connection can be made
with the work of Schmalian and Wolynes on the intrinsic instability of
classical Coulomb systems toward vastly degenerate, aperiodic stripe
phases,~\cite{SchmalianWolynes2000, SWmayo} a picture much richer than
that anticipated by Wigner almost a century
ago.~\cite{PhysRev.46.1002, TF9383400678, AndersonBasicNotions} The
present development provides what is essentially a classical density
functional theory that could be generalized to aperiodic patterns, as
in Ref.~\onlinecite{SchmalianWolynes2000}. We observe that the order
parameter must be at least two-component. Yet there are some
intrinsically quantum aspects of electronic assemblies that can {\em
  not} be done away with even if one neglects the presence of spins,
which we have largely done here except during the discussion of Wigner
solids. Indeed, interfaces between distinct CDW solutions will
generally host very special midgap electronic states that are
topological in origin~\cite{ZL_JCP, ZLMicro2} and are similar to the
solitonic midgap states in conjugated
polymers.~\cite{RevModPhys.60.781, PhysRevLett.49.1455} This important
topic has not been discussed here, to be covered in a future
submission.

Finally, we note that despite a superficial similarity between the
creation of electron-hole pairs and electron-positron pairs, the
latter does not seem to show any Urbach-like tail in the density of
states: The electron/positron mass seems to be strictly bound from
below by its apparent rest mass. Perhaps this fact can be interpreted,
in chemical terms, as the vacuum itself being composed of
``closed-shell'' units whose quantum numbers all vanish, similarly to
how noble gases have zero charge, orbital momentum, and spin, while
requiring substantial energy to open up the shell.

{\em Acknowledgments}: We thank Peter G. Wolynes, Jon C. Golden, and
Alexey Lukyanov for stimulating discussion. The financial support was
graciously provided by the NSF Grant CHE-1465125 and the Welch
Foundation Grant E-1765.

\bibliography{lowT}

\end{document}